\documentclass[8.5pt,twoside,onecolumn,american]{article}
\usepackage[super,sort&compress,comma]{natbib} 
\usepackage{mhchem}
\usepackage{times,mathptmx}
\usepackage[english]{babel}
\usepackage{sectsty}
\usepackage{balance} 

\usepackage{graphicx} %eps figures can be used instead
\usepackage{lastpage}
\usepackage[format=plain,justification=raggedright,singlelinecheck=false,font=small,labelfont=bf,labelsep=space]{caption} 
\usepackage{fancyhdr}
\usepackage[T1]{fontenc}
\usepackage[latin9]{inputenc}
\usepackage{color}
\usepackage{array}
\usepackage{float}
\usepackage{textcomp}
\usepackage{multirow}
\usepackage{amstext}
\usepackage{graphicx}
\usepackage{subscript}
\usepackage{subfig}

\begin{document}

\thispagestyle{plain}
\fancypagestyle{plain}{
\fancyhead[L]{\includegraphics[height=8pt]{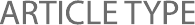}}
\fancyhead[C]{\hspace{-1cm}\includegraphics[height=20pt]{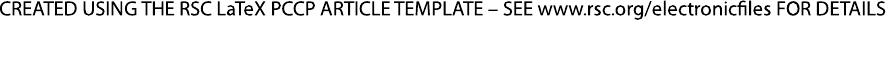}}
\fancyhead[R]{\hspace{10cm}\vspace{-0.25cm}\includegraphics[height=10pt]{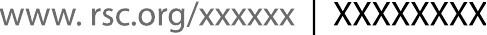}}
\renewcommand{\headrulewidth}{1pt}}
\renewcommand{\thefootnote}{\fnsymbol{footnote}}
\renewcommand\footnoterule{\vspace*{1pt}% 
\hrule width 3.4in height 0.4pt \vspace*{5pt}} 
\setcounter{secnumdepth}{5}

\makeatletter 
\renewcommand\@biblabel[1]{#1}            
\renewcommand\@makefntext[1]% 
{\noindent\makebox[0pt][r]{\@thefnmark\,}#1}
\makeatother 
\renewcommand{\figurename}{\small{Fig.}~}
\sectionfont{\large}
\subsectionfont{\normalsize} 

\fancyfoot{}
\fancyfoot[LO,RE]{\vspace{-7pt}\includegraphics[height=9pt]{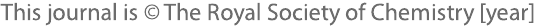}}
\fancyfoot[CO]{\vspace{-7.2pt}\hspace{12.2cm}\includegraphics{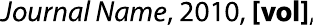}}
\fancyfoot[CE]{\vspace{-7.5pt}\hspace{-13.5cm}\includegraphics{RF}}
\fancyfoot[RO]{\footnotesize{\sffamily{1--\pageref{LastPage} ~\textbar  \hspace{2pt}\thepage}}}
\fancyfoot[LE]{\footnotesize{\sffamily{\thepage~\textbar\hspace{3.45cm} 1--\pageref{LastPage}}}}
\fancyhead{}
\renewcommand{\headrulewidth}{1pt} 
\renewcommand{\footrulewidth}{1pt}
\setlength{\arrayrulewidth}{1pt}
\setlength{\columnsep}{6.5mm}
\setlength\bibsep{1pt}

\onecolumn
  \begin{@twocolumnfalse}
\noindent\LARGE{\textbf{Tunable electronic properties of partially edge-hydrogenated armchair
boron-nitrogen-carbon nanoribbons$^\dag$}}
\vspace{0.6cm}

\noindent\large{\textbf{Naresh Alaal,\textit{$^{a,b,c}$} Nikhil Medhekar,\textit{$^{d}$} and Alok Shukla\textit{$^{b,c,\ddag}$}}}\vspace{0.5cm}
%Please note that \ast indicates the corresponding author(s) but no footnote text is required. 

\noindent\textit{\small{\textbf{Received Xth XXXXXXXXXX 20XX, Accepted Xth XXXXXXXXX 20XX\newline
First published on the web Xth XXXXXXXXXX 200X}}}

\noindent \textbf{\small{DOI: 10.1039/b000000x}}
 \end{@twocolumnfalse} \vspace{0.6cm}

\noindent  \normalsize{We employ first-principles calculations based density-functional-theory
(DFT) approach to study electronic properties of partially and fully
edge-hydrogenated armchair boron-nitrogen-carbon (BNC) nanoribbons
(ABNCNRs), with widths between 0.85 nm to 2.3 nm. Due to the partial
passivation of edges, electrons which do not participate in the bonding,
form new energy states located near the Fermi-level. Because of these
additional bands, some ABNCNRs exhibit metallic behavior, which is
quite uncommon in armchair nanoribbons. Our calculations reveal that
the metallic behavior is observed for the following passivation patterns:
(i) when B atom from one edge, and N atom from another edge, are unpassivated.
(ii) when N atoms from both the edges are unpassivated. (iii) when
C atom from one edge, and N atom from another edge, are unpassivated.
Furthermore, spin-polarization is also observed for certain passivation
schemes, which is also quite uncommon for armchair nanoribbons. Thus,
our results suggest that ABNCNRs exhibit a wide range of electronic
and {magnetic }properties in that the fully edge-hydrogenated
ABNCNRs are direct band gap semiconductors, while partially edge-hydrogenated
ones are either semiconducting, or metallic, {while
simultaneously exhibiting spin polarization,} based on the nature
of passivation. We also find that the ribbons with larger widths are
more stable, as compared to the narrower ones.}

\section*{}

\vspace{-1cm}

\footnotetext{\textit{$^{a}$IITB-Monash Research Academy, CSE Building 2$^{\,nd}$ Floor,
IIT Bombay, Mumbai 400076, India}}

\footnotetext{\textit{$^{b}$Department of Physics, Indian Institute of Technology Bombay,
Mumbai 400076, India}}

\footnotetext{\textit{$^{c}$Department of Physics, Bennett University, Plot No. 8-11,
Tech. Zone II, Greater Noida 201310 (UP) India}}

\footnotetext{\textit{$^{d}$ Department of Materials Engineering, Monash University, Clayton,
Victoria 3800, Australia}}
\footnotetext{\dag~Electronic Supplementary Information (ESI) available:See DOI: 10.1039/b000000x/}
%additional addresses can be cited as above using the lower-case letters, c, d, e... If all authors are from the same address, no letter is required

\footnotetext{\ddag~E-mail: shukla@phy.iitb.ac.in}

\section{{\normalsize{}Introduction}}

Successful synthesis of graphene\cite{graphene_synth} has led to
intensive research activities aimed at the discovery of novel nanomaterials.
Low-dimensional materials have distinctive electronic properties when
compared to their bulk counterparts, because of the effects of quantum
confinement\cite{low1,low2}. Several other two-dimensional (2D) materials
such as h-BN monolayer, transition metal dichalcogenides, phosphorene,
and borophene have been successfully synthesized over the years\cite{bn_synth,borophene,opt_tmdc,phospho}.
Graphene has received remarkable attention from the researchers due
to its unusual properties such as high carrier mobility at room temperature,
high electrical conductivity, high thermal conductivity, and excellent
mechanical strength\cite{gn1,gn2,gn3,gn4}.{{} }Although
graphene exhibits many interesting properties, but its applications
are limited as far as electronic devices are concerned, due to its
zero band gap. Thus, graphene cannot be used in those devices, which
require switching between states of low and high conductivity\cite{gn1,graphene_synth}.
Several techniques have been discovered, which can open up a band
gap in graphene, such as: (a) the use of suitable substrates, (b)
patterning into one-dimensional (1D) nanoribbons, (c) strain engineering,
(d) chemical functionalization, and (e) doping with isoelectronic
atoms\cite{bandgap_e1,bandgap_e10,bandgap_e2,bandgap_e3,bandgap_e4,bandgap_e5,bandgap_e6,bandgap_e7,bandgap_e8,bandgap_e9}. 

Another promising technique that introduces a band gap in graphene
is by forming graphene and h-BN composites. Boron-nitrogen-carbon
(BNC) monolayer synthesized by Li et al.\cite{2D_BNC}, opened the
possibility of combining graphene and 2D h-BN domains, because: (a)
C-C bond length 1.42 {\AA{} is very close to} B-N
bond length 1.44 {{} \AA , and (b) C-C and B-N units
are isoelectronic. These monolayers have completely distinct properties
from their parent materials, graphene, and 2D h-BN monolayer.}{{}
}Experimental synthesis of 2D BNC monolayer led to a large number
of theoretical, as well as experimental, studies of graphene-BN composites
consisting of CC and BN units in various proportions.\cite{bhowmick_quantum,bernardi_BNC,RF,ashwin_tunable,tunable_BNC,CBN_half,2dbnc} 

Nanoribbons, which are obtained by truncating 2D materials into 1D
ones, also exhibit interesting electronic, magnetic and optical properties
on the basis of their width, and decoration of edges\cite{nr1,nr2}.
Graphene nanoribbons (GNRs) and boron nitride nanoribbons (BNNRs)
have been experimentally fabricated by unwrapping of carbon nanotubes
and BN nanotubes, respectively\cite{gnr_synthl,bnnr_synth}. Armchair
GNRs (AGNRs) are non-magnetic semiconductors for all widths, have
oscillating band gaps over families, and approach the value of a 2D
sheet for large widths\cite{AGNR_Eg_osc}. Zigzag GNRs (ZGNRs) have
tunable band gaps from metal to semiconductor depending on the width
and passivation of edges\cite{O_zgnr,pass_zgnr}. It has also been
demonstrated theoretically that ZGNRs$ $ exhibit half-metallic behavior
when electric field is applied across their width\cite{ZGNR_Efield}.
ABNNRs are non-magnetic semiconductors, for which the band gap decreases
with increasing width of the nanoribbon. ZBNNRs are either magnetic,
or nonmagnetic, depending on their edge passivation\cite{spati_bncnr,bnnr_magnetic,half_bnnr}.

Similarly, BNC nanoribbons (BNCNRs), if synthesized, can also contain
boron, nitrogen, and carbon atoms in various proportions.
Although BNC nanoribbons have not been experimentally realized yet,
they have been studied extensively using various theoretical methods.\cite{zgnr_bn,du_jacs,kan_2008,dihydrogen,bncnr1,armchair_half,BCN_flat,spati_bncnr,Ouyang,abnc_band,bnc_transition,bnc_zig,Fan,bcn1,bcn2}.
Huang et al.\cite{bncnr1} studied the hybrid bare BNC nanoribbons
obtained by combining BNNRs and GNRs, by using first-principles DFT
approach. {They found that armchair BNC nanoribbons
(ABNCNRs) are non-magnetic semiconductors, while zigzag BNC nanoribbons
(ZBNCNRs) exhibit half-metallicity for certain widths, and C/BN compositions.}
Liu et al.\cite{armchair_half} studied the electronic structures
of hybrid ABNCNRs by using the DFT approach. They found that these
nanoribbons are magnetic metals when the B and N atoms {are}
unpaired, and that they also exhibit half-metallic behavior when the
O atoms {are} adsorbed on appropriate positions of
nanoribbons. Kan et al.\cite{kan_2008} studied spin-polarized electronic
properties of fully hydrogen-passivated zigzag BNC nanoribbons by
using first-principles DFT, and reported that due to a competition
between charge and spin polarization, these nanoribbons can exhibit
half-metallic behavior. Fan et al.\cite{Fan} studied unpassivated
ZBNCNR structures obtained by joining ZGNRs and ZBNNRs along the width,
and they found that the electronic properties can be tuned by changing
the width of the ZGNR domain. Dihydrogenated ZBNCNRs were studied
by Liu \emph{et al}. \cite{dihydrogen}, and they reported that the
half-metallic property depends on the width of both the carbon, and
the BN domains. Basheer \emph{et al.}\cite{spati_bncnr} studied fully
hydrogen-passivated, and partially passivated ZBNCNRs, which are composed
of an equal number of C, B and N atoms. They observed that the half-metallic
behavior depends on the edge passivating atoms, and the width of the
ribbon. 

Generally, edge passivation plays a key role in modifying electronic
structures of nanoribbons \cite{GNR_GW,bn_pass,phospho_pass,Nalaal1,Nalaal2}.
Interestingly, partial edge passivation also leads to interesting
electronic properties in nanoribbons\cite{half_bare_zig,half_bn}.
For example, ABNNRs were found to be half-metallic when edge B atoms
are passivated, and edge N atoms are left unpassivated\cite{half_bn}.
Similarly, zigzag SiC naniribbons behave as magnetic metals when edge
Si atoms are unpassivated, while they become magnetic semiconductors,
when edge C atoms are unpassivated\cite{half_bare_zig}. Although
one study of partially passivated ZBNCNRs exists\cite{spati_bncnr},
to the best of our knowledge, no such studies of ABNCNRs have been
undertaken.

In this work, we present a detailed study of electronic properties
of ABNCNRs, whose edges are fully and partially passivated with hydrogen
atoms. Our calculations suggest that by partial edge passivation,
one can tune electronic properties of ABNCNRs in a variety of ways
such as non-magnetic semiconductors, magnetic semiconductors, or magnetic
metals, depending on the nature of edges, and their passivation. This
is an interesting result because normally no magnetic behavior is
expected in case of armchair nanoribbons. We also found that few configurations
of ABNCNRs exhibit metallic behavior irrespective of their widths,
and ribbons with larger width are more stable as compared to the narrower
ones. 

Remainder of the paper is organized as follows. In the next section
we describe our theoretical methodology, in section III we present
and discuss our results, while in section IV we summarize and conclude.

\section{{\normalsize{}Computational Details}}

\label{sec:theory}

{Computational methodology followed in this work
is similar to the one adopted in our previous works on SiCNRs,\cite{Nalaal1,Nalaal2}
except that the present calculations are limited to the level of density-functional
theory (DFT). We performed all the calculations using the software
package VASP\cite{VASP}, in which the ground state properties are
computed using a plane-wave based first-principles, DFT approach.
We chose to adopt the generalized-gradient approximation (GGA) approach
for the exchange correlation functional, coupled with Perdew, Burke,
Enzerhof (PBE) pseudopotentials.\cite{psp,pbe}} A kinetic energy
cut-off of 500 eV was used, along with a k-point grid of $11\times1\times1$,
employed for structural relaxation.{{} Convergence
threshold energy of $10^{-4}$ eV was used, and atoms were relaxed
until the force on each atom was less than $0.01$ eV/\AA . After
the geometry optimization,} charge density profiles were computed
using a superior k-point mesh of $45\times1\times1$. Nanoribbons
studied here are considered periodic in the $x-$direction, and vacuum
(intercell distance) of more than 14 {\AA{}} has
been taken in $y$ and $z$ directions to model this 1D structure.
The widths of ABNCNRs are represented by the integer $N_{a}$, which
equal to the total number of C-C/BN pairs across the width. In this
work we present our calculations on ABNCNRs with widths ranging between
0.85 nm to 2.3 nm, corresponding to ribbons with $6\leq N_{a}\leq18$.

\section{{\normalsize{}Results And Discussion}}

\label{sec:results}

\subsection{Band Structures of Graphene and Boro-Nitride Nanoribbons}

Before considering the cases of partially saturated ABNCNRs, we compute
the band structures of armchair type graphene nanoribbons (AGNRs),
and boron nitride nanoribbons (ABNNRs). 

\begin{figure}[H]
\includegraphics[scale=0.33]{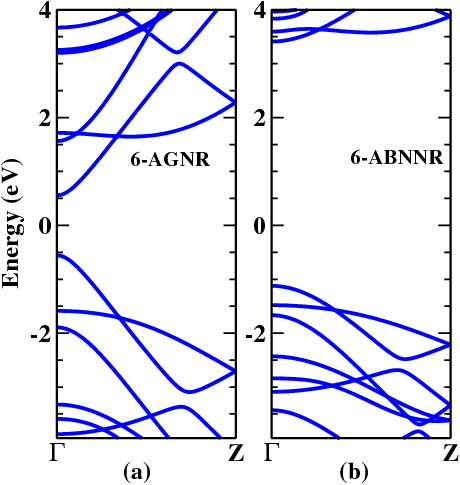}

\caption{Band Structures of (a) 6-AGNR and (b) 6-ABNNR}

\label{band_gb}
\end{figure}

Because ABNCNRs of a given width are combinations of AGNRs and ABNNRs,
therefore, in order to achieve a deeper understanding of their electronic
structure, it will be helpful to know the band structures of the original,
unmixed nanoribbons. In particular, we compute the band structures
of 6-AGNR and 6-ABNNR, and in both the cases we assume that the edge
atoms are saturated by hydrogens. The calculated band structures of
these ribbons, along the line connecting high symmetry Brillioun zone
points, $\Gamma$ and X, are presented in Fig. \ref{band_gb}. As
is obvious from the figure that for both the ribbons, the valence
band maximum (VBM), and the conduction band minimum (CBM) occur at
high symmetry Brillioun zone point $\Gamma$, implying that these
are direct band gap materials. The calculated band gaps of 1.11 eV
(6-AGNR) and 4.53 eV (6-ABNNR) are in good agreement with the results
reported by other authors\cite{GNR_GW4,S_gnr}. This establishes the
accuracy of the computational methodology employed by us, which we
utilize in the following sections to study the electronic structure
of various configurations of ABNCNRs.

\begin{figure}[H]
\begin{centering}
\includegraphics[scale=0.8]{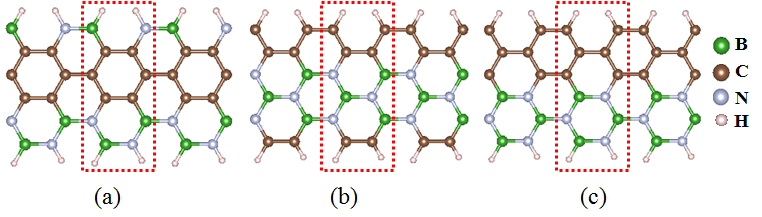}
\par\end{centering}
\caption{Geometrical structures of fully hydrogen saturated (a) BN-BN-ABNCNR,
(b) CC-CC-ABNCNR, (c) CC-BN-ABNCNR. Unit cells considered for the
calcuations are shown in red dashed boxes.}

\label{str_abnc}
\end{figure}

\subsection{Nature of Bonds and Stability of ABNCNRs}

Henceforth we consider ABNCNRs, which consist of an equal number of
C-C and B-N dimers, along the width. We use the notation x-y-ABNCNR
to denote a given nanoribbon, in which x and y denote the dimer units
terminating the two edges. Thus, based upon their edge terminations,
nanoribbons are divided into three configurations labeled as BN-BN-ABNCNR,
CC-CC-ABNCNR, and CC-BN-ABNCNRs. In BN-BN-, and CC-CC-ABNCNRs, both
the edges are composed {only of BN and CC dimer units,
respectively.} In the CC-BN configuration, one edge is composed of
CC dimers, while the other one is composed of BN dimers. Geometrical
structures of ABNCNRs are presented in Fig. \ref{str_abnc}. From
previous studies\cite{2D_BNC,akmanna1,spati_bncnr}, 2D BNC nanostructures
which consist of maximum number of B-N and C-C bonds, are more stable
as compared to any other structures, which have other types of bonds
(say, B-C, N-C, B-B and N-N, bonds). Here we quantitatively examine
the stability of various configurations of ABNCNRs, as a function
of the number of bonds of various types present in the nanoribbon,
and present the results in Table \ref{tab_bonds}. All the nanoribbons
considered are assumed to have hydrogen passivated edges. We do not
consider structures with B-B and N-N bonds in our study, because they
are even more unstable\cite{spati_bncnr,akmanna2}. 

\begin{table}
\begin{tabular}{|c|c|c|c|c|c|}
\hline 
Structure & No. of  & No. of  & No. of  & No. of  & {Relative }\tabularnewline
 & B-N bonds & C-C bonds & B-C bonds & N-C bonds & {energy (eV)}\tabularnewline
\hline 
\hline 
6-BN-BN-ABNCNR & 6 & 3 & 2 & 2 & 1.38\tabularnewline
\hline 
6-CC-CC-ABNCNR & 5 & 4 & 2 & 2 & 0.63\tabularnewline
\hline 
6-CC-BN-ABNCNR & 6 & 5 & 1 & 1 & 0\tabularnewline
\hline 
\end{tabular}

\caption{Number of various types of bonds present in the 6-BN-BN-ABNCNR, 6-CC-CC-ABNCNR
and 6-CC-BN-ABNCNR, and their total energies/cell.\label{tab_bonds}}
\end{table}

From Fig. \ref{str_abnc} and Table \ref{tab_bonds}, 6-CC-BN-ABNCNR
consists of maximum number of C-C (5) and B-N (6) bonds, and it is
clear that the total energies of 6-BN-BN-ABNCNR and 6-CC-CC-ABNCNR
are 1.38 eV and 0.63 eV, respectively, higher than that of 6-CC-BN-ABNCNR.
Total energies per unit cell of these nanoribbons are in following
order: CC-BN-ABNCNR<CC-CC-ABNCNR<BN-BN-ABNCNR. Thus, CC-BN-ABNCNR
corresponds to the most stable configuration, when compared with the
other two types of ABNCNRs. We conclude that these nanoribbons become
more unstable with the increasing number of B-C and N-C bonds, and
with the decreasing numbers of C-C and B-N bonds.

\subsection{Electronic properties of BN-BN-ABNCNRS}

\begin{figure}[H]
\begin{centering}
\includegraphics[scale=0.75]{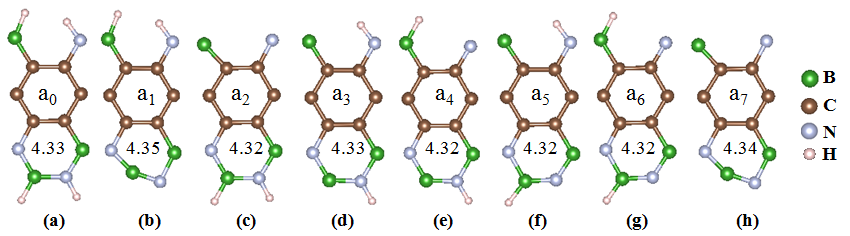}
\par\end{centering}
\caption{Geometric structures of eight configurations of BN-BN-ABNCNRs unit
cells: (a) B\protect\textsuperscript{H}N\protect\textsuperscript{H}-B\protect\textsuperscript{H}N\protect\textsuperscript{H},
(b) B\protect\textsuperscript{H}N\protect\textsuperscript{H}-BN,
(c) BN-B\protect\textsuperscript{H}N\protect\textsuperscript{H},
(d) BN\protect\textsuperscript{H}-BN\protect\textsuperscript{H},
(e) B\protect\textsuperscript{H}N\protect\textsuperscript{}-BN\protect\textsuperscript{H},
(f) BN\protect\textsuperscript{H}-B\protect\textsuperscript{H}N,
(g) B\protect\textsuperscript{H}N\protect\textsuperscript{}-B\protect\textsuperscript{H}N,
(h) BN-BN. Superscript H on a given edge atom indicates its saturation
by hydrogen. For the sake of brevity, we represent these configurations
by notations a\protect\textsubscript{0} through a\protect\textsubscript{7},
as shown in the figure. {Lattice constants for all
the configurations are displayed in \AA{} units.}}

\label{Fig:str_bn}
\end{figure}

BN-BN-ABNCNRs are divided into eight configurations based on how the
edge atoms are passivated with the hydrogen atoms. These configurations
can be described as: (a) both the upper and lower edge atoms are passivated
with hydrogen, (b) the upper edge atoms are passivated with H, and
lower edge atoms are unpassivated, (c) lower edge atoms are passivated
with H, and upper edge atoms are unpassivated, (d) N atoms from both
the edges are terminated with H, while B atoms from both the edges
are unpassivated, (e) B atoms from both the edges are terminated with
H, while N atoms from both the edges are unpassivated, {(f)
B atoms on the upper edges and N atoms on the lower edge lower are
passivated with H, while remaining edge atoms are unpassivated, (g)
N atoms on the upper edge and B atoms on the lower edge are passivated
with H, while remaining edge atoms are unpassivated.} (h) Both the
edges are unpassivated. These configurations are labeled as: B\textsuperscript{H}N\textsuperscript{H}-B\textsuperscript{H}N\textsuperscript{H}-ABNCNR
(a\textsubscript{0}), B\textsuperscript{H}N\textsuperscript{H}-BN-ABNCNR
(a\textsubscript{1}), BN-B\textsuperscript{H}N\textsuperscript{H}-ABNCNR
(a\textsubscript{2}), BN\textsuperscript{H}-BN\textsuperscript{H}-ABNCNR
(a\textsubscript{3}), B\textsuperscript{H}N\textsuperscript{}-BN\textsuperscript{H}-ABNCNR
(a\textsubscript{4}), BN\textsuperscript{H}-B\textsuperscript{H}N-ABNCNR
(a\textsubscript{5}), B\textsuperscript{H}N-B\textsuperscript{H}N-ABNCNR
(a\textsubscript{6}), BN-BN-ABNCNR (a\textsubscript{7}). Thus, in
this notation, superscript H on an edge atom indicates its passivation
by hydrogen. Fig. \ref{Fig:str_bn} displays the geometric structures
and {lattice parameters} of these eight configurations
of BN-BN-ABNCNRs, corresponding to various edge-passivation schemes.
Configuration a\textsubscript{0}, and a\textsubscript{7}, represent
fully hydrogen-passivated, and bare BN-BN-ABNCNRs, respectively. {These
ABNCNRs have different lattice constants ranging from 4.32 to 4.35
\AA{}.}

\begin{figure}[H]
\begin{centering}
\includegraphics[scale=0.8]{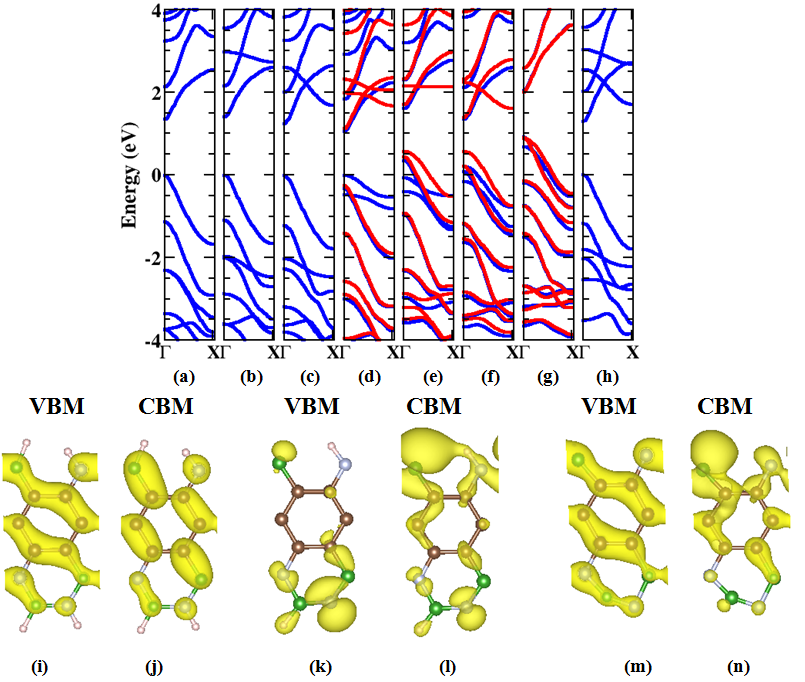}
\par\end{centering}
\caption{Band structures of configurations (a) a\protect\textsubscript{0},
(b) a\protect\textsubscript{1}, (c) a\protect\textsubscript{2},
(d) a\protect\textsubscript{3}, (e) a\protect\textsubscript{4},
(f) a\protect\textsubscript{5}, (g) a\protect\textsubscript{6},
and (h) a\protect\textsubscript{7}. Partial charge density profiles:
(i) and (j) VBM and CBM of a\protect\textsubscript{0}, (k) and (l)
VBM and CBM of a\protect\textsubscript{5}, (m) and (n) VBM and CBM
of a\protect\textsubscript{7}. Blue and red curves denote bands corresponding
to opposite spin orientations. Colors of atoms are consistent with
Fig. \ref{Fig:str_bn}.}

\label{Fig:band_bn}
\end{figure}

Band structures for all the BN-BN-ABNCNRs configurations are presented
in Fig. \ref{Fig:band_bn}, while Table \ref{table1} presents the
band gaps, {magnetic moments, energy difference $(\Delta E$
) between magnetic and nonmagnetic states, and the values of Gibbs
free energy of formation ($\delta G$) for all the configurations}
of BN-BN-ABNCNRs. All the band structures are plotted along the line
joining high symmetry Brillioun zone points $\Gamma$ and $X$, while
the Fermi-level has been set to 0 eV. Fig. \ref{Fig:band_bn} shows
that configurations a\textsubscript{0}, a\textsubscript{1}, a\textsubscript{2},
and a\textsubscript{7 }are direct band gap semiconductors because
VBM and CBM for these structures occur at same high Brillioun zone
point $\Gamma$. In Figs. \ref{Fig:band_bn} (d)-(g), blue lines and
red lines represent band structures for spin-up and spin-down channels,
respectively. Thus, it is obvious that configurations a\textsubscript{3},
a\textsubscript{4}, a\textsubscript{5}, and a\textsubscript{6}
exhibit spin-polarized behavior, which would not have been possible
for fully hydrogen-passivated ABNCNRs. Furthermore, from Table \ref{table1}
it is clear that, configuration a\textsubscript{3} is a magnetic
semiconductor because band gaps for up and down spins are unequal,
and in the semiconducting range. Examination of Fig. \ref{Fig:band_bn}
and Table \ref{table1} further reveals that configurations a\textsubscript{4},
a\textsubscript{5}, and a\textsubscript{6} are magnetic metals because:
(a) their bands are spin-polarized, (b) valence and conduction bands
for both spin orientations cross the Fermi-level, and (c) they have
finite magnetic moment per unit cell.{{} }Thus, in partially
edge-hydrogenated BN-BN-ABNCNRs, spin-polarization is observed only
when one atom from the upper edge, and another one from the lower
edge, are unpassivated. In particular, metallic behavior appears when:
(a) B atom from one edge, and N atom from another one, are passivated,
or (b) when N atoms from both the edges are unpassivated.

\begin{table}
\begin{centering}
\begin{tabular}{|c|c|c|c|c|c|c|c|c|}
\hline 
\multirow{2}{*}{Configuration} & \multicolumn{2}{c|}{Band gap (eV) } & \multirow{2}{*}{$MM$ ($\mu_{B}$)} & \multirow{2}{*}{{Type}} & \multirow{2}{*}{{$\Delta E$ (eV)}} & \multirow{2}{*}{{$\delta G$ (eV)}} & \multirow{2}{*}{{$E_{f}$ (eV/\AA )}} & \multirow{2}{*}{{$E_{b}$ (eV)}}\tabularnewline
\cline{2-3} 
 & Up  & Down &  &  &  &  &  & \tabularnewline
\hline 
a\textsubscript{0} & 1.34 & 1.34 & {0} & {NM} & - & {0.180} & {0.332 } & {-4.803}\tabularnewline
\hline 
a\textsubscript{1} & 1.40 & 1.40 & {0} & {NM} & - & {0.411} & {0.660} & {-4.821}\tabularnewline
\hline 
a\textsubscript{2} & 1.24 & 1.24 & {0} & {NM} & - & {0.413} & {0.669} & {-4.803}\tabularnewline
\hline 
a\textsubscript{3} & 1.04 & 1.37 & {0.98} & {FM} & {0.894} & {0.536} & {0.868} & {-3.944}\tabularnewline
\hline 
a\textsubscript{4} & M & M & {0.92} & {FM} & {0.519} & {0.539} & {0.873} & {-3.922}\tabularnewline
\hline 
a\textsubscript{5} & M & M & {0.01} & {Ferri} & {0.314} & {0.554} & {0.898} & {-3.814}\tabularnewline
\hline 
a\textsubscript{6} & M & M & {0.12} & {Ferri} & {0.004} & {0.550} & {0.891} & {-3.845}\tabularnewline
\hline 
a\textsubscript{7} & 1.30 & 1.30 & {0} & {NM} & - & {0.724} & {0.999} & {-}\tabularnewline
\hline 
\end{tabular}
\par\end{centering}
\caption{Band gaps for the two spin orientations, magnetic moments per unit
cell ($MM)$, {type of magnetic behavior, energy difference
between magnetic and nonmagnetic states ($\Delta E$), Gibbs free
energy of formation per atom ($\delta G$), edge formation energy
per unit length ($E_{f}$), and binding energy per H atom ($E_{b}$)
of various configurations of BN-BN-ABNCNRs. NM, FM, and Ferri represent
nonmagnetic, ferromagnetic and ferrimagnetic behaviors, respectively.
M in the band gap column implies metallic behavior. }\label{table1}}
\end{table}

\begin{figure}[H]
\includegraphics{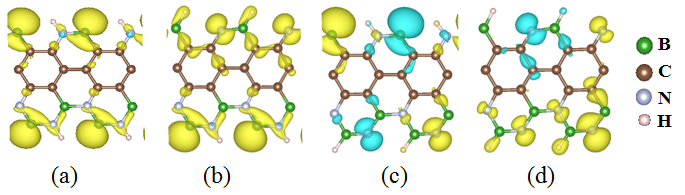}

\caption{{Spin density plots for (a) a$_{3}$, (b) a$_{4}$,
(c) a$_{5}$, and (d) a$_{6}$ configurations. Yellow and blue colors
represent spin-up, and spin-down, states, respectively}}

\label{spin_bn}
\end{figure}

{To understand the atomic contributions to the magnetic
moment of the unit cell, we have performed calculations for the supercells
containing double units and presented spin density plots of the four
configurations (a\textsubscript{3}-a\textsubscript{6}) in Fig. \ref{spin_bn}.
For the configurations a$_{3}$-a$_{4}$, we note that the magnetic
moments are predominantly due to the spin-up states, derived mainly
from the unpassivated edge atoms, and have large magnetic moments
with ferromagnetic behavior. In case of a$_{5}$ and a$_{6}$, we
find that the magnetic moments are too small when compared with other
spin-polarized configurations (see Table \ref{table1}), because spins
on the two edges are almost equal in magnitude, and oppositely oriented,
resulting in a small net magnetic moment. Thus, both these configurations
are ferrimagnetic metals. The energy difference $\Delta E$ between
the magnetic and nonmagnetic states tells that a\textsubscript{{3}}
is the most stable magnetic configuration, and a\textsubscript{{6}}
is the least stable one.} Thus, we conclude that partially passivated
BN-BN ABNCNRs can be non-magnetic semiconductors, magnetic semiconductors,
or magnetic metals, depending upon the nature of edge passivation
by H atoms.

We also calculated the partial charge density profiles in order to
understand the contribution of various atoms to the valence band maxima
(VBM) and the conduction band minima (CBM) of BN-BN-ABNCNRs, and in
Figs. \ref{Fig:band_bn} (i)\textendash (m) we present those for configurations
$\text{a}_{\text{0}}$, $\text{a}_{5}$, and a\textsubscript{7}.
We observe that, in configuration $\text{\ensuremath{a_{0}}}$, the
VBM and CBM derive their dominant contribution from the C-C pairs,
with upper edge atoms also contributing significantly. In the partially
passivated a\textsubscript{5} configuration, the VBM and CBM are
mainly localized on unpassivated atoms. In configuration a\textsubscript{7},
we can observe that the edge atoms undergo reconstruction, and that
the VBM is mainly composed of contributions from C-C pairs, while
the CBM is localized on the unpassivated upper edge atoms. 

\begin{figure}[H]
\includegraphics[scale=0.5]{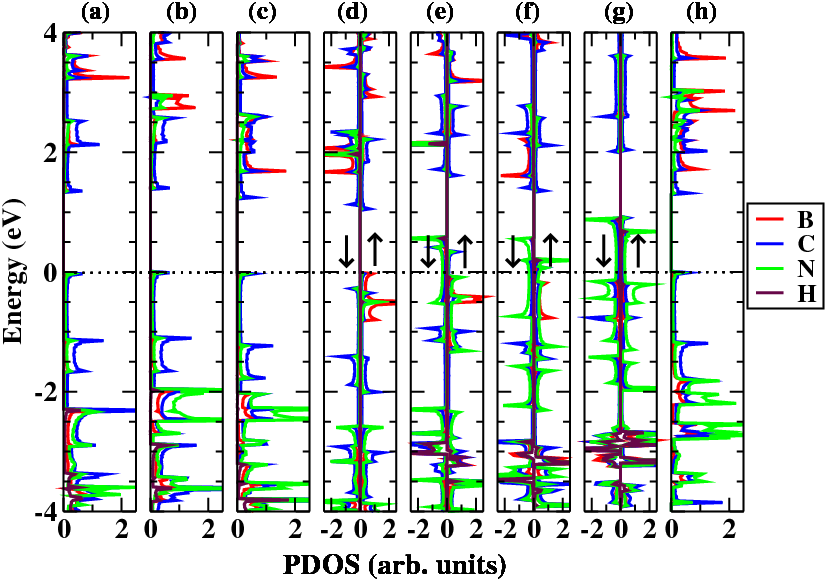}

\caption{Projected density of states of atoms B, C, N and H for all the configurations
of 6-BN-BN-ABNCNR: (a) a\protect\textsubscript{0}, (b) a\protect\textsubscript{1},
(c) a\protect\textsubscript{2}, (d) a\protect\textsubscript{3},
(e) a\protect\textsubscript{4}, (f) a\protect\textsubscript{5},
(g) a\protect\textsubscript{6} and (h) a\protect\textsubscript{7}.
Up and down arrows indicate spin-up and spin-down, states, respectively.
Fermi level has been set at 0 eV, and represented by a black dashed
line. \label{pdos_bn}}
\end{figure}

\begin{figure}[H]
\includegraphics[scale=0.5]{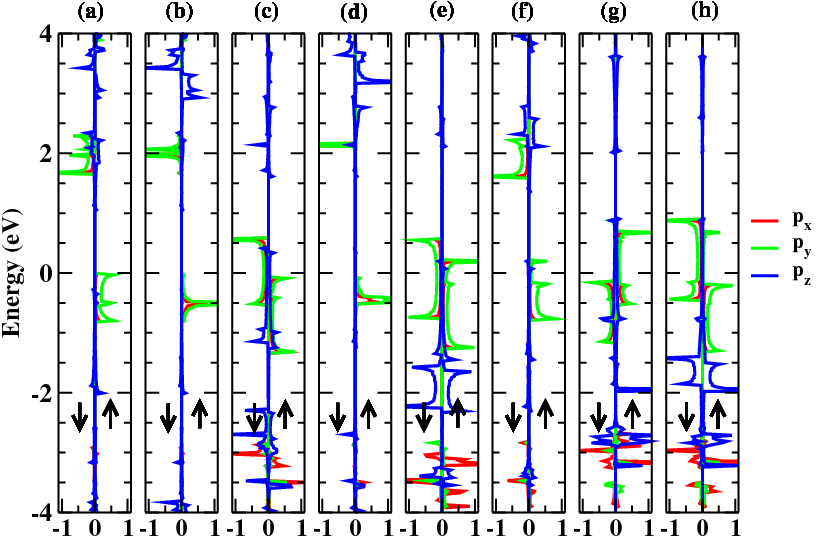}

\caption{PDOS of $p$ orbitals of atoms: (a) upper edge bare B atom of a\protect\textsubscript{3},
(b) lower edge bare B atom of a\protect\textsubscript{3}, (c) upper
edge bare N atom of a\protect\textsubscript{4}, (d) lower edge bare
B atom of a\protect\textsubscript{4}, (e) upper edge bare N atom
of a\protect\textsubscript{5}, (f) lower edge bare B atom of a\protect\textsubscript{5},
(g) upper edge bare N atom of a\protect\textsubscript{6}, (h) lower
edge bare N atom of a6. Up and down arrows denote spin-up, and spin-down,
states, respectively. \label{pdos_orb_bn}}
\end{figure}

In order to understand the origins of changes in the band structures
of these nanoribbons, with various edge hydrogenation patterns, we
present projected density of states (PDOS) for individual atoms in
the Fig. \ref{pdos_bn}. We also present PDOS corresponding to the
$p$ orbitals of the bare edge atoms of spin-polarized configurations
in the Fig. \ref{pdos_orb_bn}. We find that in the nonmagnetic BN-BN-ABNCNRs
(a\textsubscript{0}, a\textsubscript{1}, a\textsubscript{2} and
a\textsubscript{7}), VBM and CBM originate mainly from the C atoms,
with small contribution from N atoms. In all the configurations, H
atoms contribute to the states which are away from the Fermi level
in the valence band. In the spin-up channel of configuration a$_{3}$,
VBM originates mainly from the B atoms, and particularly from \textit{p\textsubscript{\textit{y}}}
orbital of upper edge B atom, while CBM is derived from C atoms (See
Fig. \ref{pdos_bn} (d) and Figs.\ref{pdos_orb_bn}(a)-(b)). In the
spin-down channel, VBM and CBM both originate from C atoms. In the
spin-up and spin-down channels of configurations of a\textsubscript{4}-a\textsubscript{6},
VBMs cross Fermi-level, because of the additional energy states that
are mainly contributed by the $p_{y}$ orbitals of the N atoms (See
Figs. \ref{pdos_bn} (e)-(g) and Figs. (c)-(h)), causing them to become
metallic.

{We also performed Bader charge analysis \cite{bader}
for various nanoribbons, and in Table \ref{bader_bn}
we present the computed Bader charges on the edge atoms of the spin-polarized
BN-BN configurations a\textsubscript{{3}}-a\textsubscript{{6}}
, and compare them with the charges on edge atoms of the bare BN-BN-ABNCNR
(configuration a\textsubscript{{7}}). From the table
we observe that in the configurations a\textsubscript{{3}}-a\textsubscript{{5}},
bare edge atoms accumulate the electrons, while the hydrogenated ones
lose them. For example, in case of a\textsubscript{{3}},
the left upper edge (LUE) atom (B) and left lower edge (LLE) (B) atom
gain the charge of 0.46 $e$ and 0.41 $e$ respectively. On the other
hand, right upper edge (RUE) atom (N) and right lower edge (RLE) atom
(N) lose the charges of 0.24 $e$, and 0.06 $e$, respectively. In
configuration a\textsubscript{{6}}, one bare edge
accumulates the electronic charge, while the other one loses it. One
common feature is when hydrogen atoms passivate the boron atoms (H
on LUE and H on LLE) they show high electronegativity, as compared
with nitrogens passivated by hydrogen atoms (H on RUE and H on RLE).}

\begin{table}
\centering{}%
\begin{tabular}{|c|c|c|c|c|c|c|c|c|}
\hline 
\multirow{2}{*}{{Configuration}} & \multirow{2}{*}{{LUE (B) }} & \multirow{2}{*}{{RUE(N)}} & \multirow{2}{*}{{LLE (B)}} & \multirow{2}{*}{{RLE (N)}} & \multirow{2}{*}{{H-LUE}} & \multirow{2}{*}{{H-RUE}} & \multirow{2}{*}{{H-LLE}} & \multirow{2}{*}{{H-RLE}}\tabularnewline
 &  &  &  &  &  &  &  & \tabularnewline
\hline 
{a\textsubscript{3}} & {1.87 (+0.46)} & {6.17(-0.24)} & {1.75 (+0.41)} & {6.63(-0.06)} & - & {0.676} & - & {0.619}\tabularnewline
\hline 
{a\textsubscript{4}} & {1.40(-0.01)} & {6.11(+0.30)} & {1.71(+0.37)} & {6.62(-0.07)} & {1.585} & - & - & {0.680}\tabularnewline
\hline 
{a\textsubscript{5}} & {1.83(+0.42)} & {6.21(-0.20)} & {0.90(-0.44)} & {6.76(+0.07)} & - & {0.683} & {1.639} & -\tabularnewline
\hline 
{a\textsubscript{6}} & {1.43(+0.02)} & {6.01(-0.40)} & {0.91(-0.43)} & {6.76(+0.07)} & {1.537} & - & {1.636} & -\tabularnewline
\hline 
{a\textsubscript{7}} & {1.41} & {6.41} & {1.34} & {6.69} & - & - & - & -\tabularnewline
\hline 
\end{tabular}\caption{{Total Bader charges on the edge atoms and hydrogen
atoms of configurations of a\protect\textsubscript{{3}}-a\protect\textsubscript{{7}}.
Positive and negative values denote the electron gain and loss, respectively.
The values given in the parentheses show the charge difference with
respect to the bare (a\protect\textsubscript{{7}})
configuration. Above, acronyms LUE, RUE, LLE, and RLE stand for ``left
upper edge'', ``right upper edge'', ``left lower edge'', and
``right lower edge'', respectively. Notations such as H-LUE denote
hydrogen atom bonded with the LUE etc. All the charges are in electron
units $e$ ( 1$e$ = 1.62x10$^{-19}$C).}\label{bader_bn}}
\end{table}

\subsection{Electronic properties of CC-CC-ABNCNRS$ $}

\begin{figure}[H]
\begin{centering}
\includegraphics[scale=0.75]{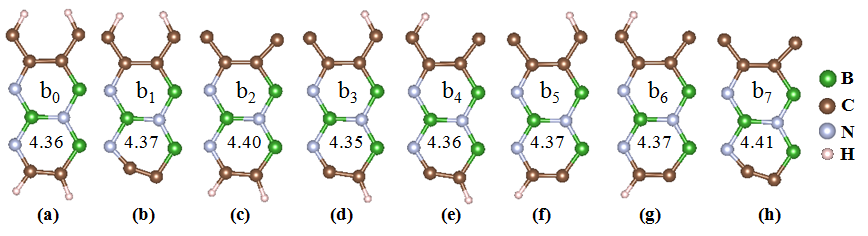}
\par\end{centering}
\caption{Geometric structures of eight configurations of CC-CC-ABNCNRs unit
cells: (a)C\protect\textsuperscript{H}C\protect\textsuperscript{H}-C\protect\textsuperscript{H}C\protect\textsuperscript{H}
, (b) C\protect\textsuperscript{H}C\protect\textsuperscript{H}-CC,
(c) CC-C\protect\textsuperscript{H}C\protect\textsuperscript{H},
(d) CC\protect\textsuperscript{H}-CC\protect\textsuperscript{H},
(e) C\protect\textsuperscript{H}C\protect\textsuperscript{}-CC\protect\textsuperscript{H},
(f) CC\protect\textsuperscript{H}-C\protect\textsuperscript{H}C,
(g) C\protect\textsuperscript{H}C-C\protect\textsuperscript{H}C
, (h) CC-CC. For the sake of brevity, we represent these configurations
by notations b\protect\textsubscript{0} through b\protect\textsubscript{7},
as shown in the figure. {Lattice constants for all
the configurations are displayed in \AA{} units.}}

\label{Fig:str_cc}
\end{figure}

CC-CC-ABNCNRs are also divided into eight configurations based on
how their edge atoms are passivated with hydrogen atoms. Similar to
BN-BN BNCNRs, CC-CC-ABNCNRs are also divided into eight configurations,
based on their edge passivation patterns, as presented in Fig. \ref{Fig:str_cc}:
C\textsuperscript{H}C\textsuperscript{H}-C\textsuperscript{H}C\textsuperscript{H}-ABNCNR
(b\textsubscript{0}), C\textsuperscript{H}C\textsuperscript{H}-CC-ABNCNR
(b\textsubscript{1}), CC-C\textsuperscript{H}C\textsuperscript{H}-ABNCNR
(b\textsubscript{2}), CC\textsuperscript{H}-CC\textsuperscript{H}-ABNCNR
(b\textsubscript{3}), C\textsuperscript{H}C-CC\textsuperscript{H}-ABNCNR
(b\textsubscript{4}), CC\textsuperscript{H}-C\textsuperscript{H}C\textsuperscript{}-ABNCNR
(b\textsubscript{5}),\textbf{ }C\textsuperscript{H}C-C\textsuperscript{H}C-ABNCNR
(b\textsubscript{6}), CC-CC-ABNCNR (b\textsubscript{7}). As is obvious
from Fig. \ref{Fig:str_cc}, configurations b\textsubscript{0}and
b\textsubscript{7} represent fully hydrogen-passivated, and completely
bare CC-CC-ABNCNRs, respectively. {These ABNCNRs have
lattice constants in the range 4.35 to 4.41 \AA{} (See Fig.
\ref{Fig:str_cc}).}

\begin{figure}[H]
\includegraphics[scale=0.8]{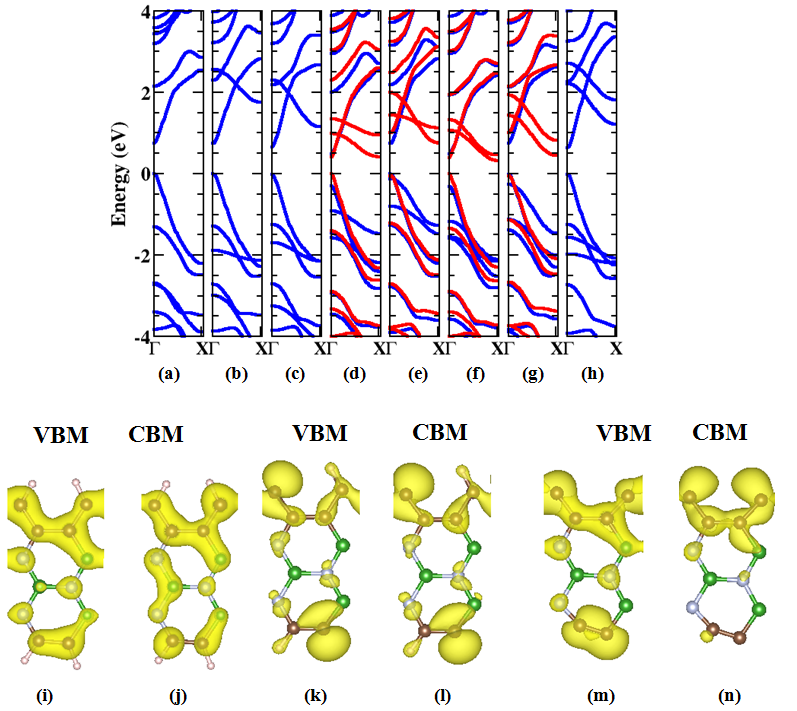}

\caption{Band structures of (a) b\protect\textsubscript{0}, (b) b\protect\textsubscript{1},
(c) b\protect\textsubscript{2}, (d) b\protect\textsubscript{3},
(e) b\protect\textsubscript{4}, (f) b\protect\textsubscript{5},
(g) b\protect\textsubscript{6}, and (h) b\protect\textsubscript{7}
configurations. Partial charge density profiles: (i) and (j) VBM and
CBM of b\protect\textsubscript{0}, (k) and (l) VBM and CBM of b\protect\textsubscript{5},
(m) and (n) VBM and CBM of b\protect\textsubscript{7}. Blue and red
curves denote bands corresponding to opposite spin orientations. Colors
of atoms are consistent with Fig. \ref{Fig:str_cc}.}

\label{Fig:band_cc}
\end{figure}

In Fig. \ref{Fig:band_cc} we present the band structures of all the
configurations of CC-CC-ABNCNRs, while Table \ref{tabl2} contains
corresponding band gaps, magnetic moments, and cohesive energies.
{From Figs. }\ref{Fig:band_cc}{{}
(a)-(c), and (h) }it is obvious that the nanoribbons with configurations
b\textsubscript{0}, b\textsubscript{1}, b\textsubscript{2}, and
b\textsubscript{7}, are all direct band gap semiconductors. Band
structures of configurations b\textsubscript{3}, b\textsubscript{4},
b\textsubscript{5}, and b\textsubscript{6} presented in Figs.\ref{Fig:band_cc}
(d)-(g), and band gaps and magnetic moments presented in Table \ref{tabl2},
make it obvious that these nanoribbons exhibit spin-polarized behavior,
with different band gaps for the two spin orientations, both of which
are in the semiconducting range. Thus, we conclude that CC-CC-ABNCNRs
configurations are either magnetic, or non-magnetic semiconductors,
but never metallic.

\begin{table}
\begin{centering}
\begin{tabular}{|c|c|c|c|c|c|c|c|c|}
\hline 
\multirow{2}{*}{Configuration} & \multicolumn{2}{c|}{Band gap (eV) } & \multirow{2}{*}{{$MM$($\mu_{B}$)}} & \multirow{2}{*}{{Type}} & \multirow{2}{*}{{$\Delta E$ (eV)}} & \multirow{2}{*}{{$\delta G$ (eV)}} & \multirow{2}{*}{{$E_{f}$ (eV/\AA )}} & \multirow{2}{*}{{$E_{b}$ (eV)}}\tabularnewline
\cline{2-3} 
 & Up  & Down &  &  &  &  &  & \tabularnewline
\hline 
b\textsubscript{0} & 0.75 & 0.75 & {0} & {NM} & - & {0.133 } & {0.244} & {-5.255}\tabularnewline
\hline 
b\textsubscript{1} & 0.74 & 0.74 & {0} & {NM} & - & {0.417} & {0.667} & {-5.305}\tabularnewline
\hline 
b\textsubscript{2} & 0.65 & 0.65 & {0} & {NM} & - & {0.429} & {0.683} & {-5.215}\tabularnewline
\hline 
b\textsubscript{3} & 0.78 & 0.40 & {0} & {AFM} & {0.687 } & {0.517} & {0.832} & {-4.601}\tabularnewline
\hline 
b\textsubscript{4} & 0.75 & 0.74 & {0} & {AFM} & {0.074} & {0.504} & {0.809} & {-4.694}\tabularnewline
\hline 
b\textsubscript{5} & 0.63 & 0.32 & {0} & {AFM} & {0.054} & {0.530} & {0.851} & {-4.513}\tabularnewline
\hline 
b\textsubscript{6} & 0.79 & 0.41 & {0} & {AFM} & {0.105} & {0.515} & {0.828} & {-4.614}\tabularnewline
\hline 
b\textsubscript{7} & 0.64 & 0.64 & {0} & {NM} & - & {0.812} & {1.105} & {-}\tabularnewline
\hline 
\end{tabular}
\par\end{centering}
\caption{Band gaps for the two spin orientations, magnetic moments per unit
cell ($MM)$, {type of magnetic behavior, energy difference
between magnetic and nonmagnetic states, Gibbs free energy of formation
per atom ($\delta G$), edge formation energy per unit length ($E_{f}$),
and binding energy per H atom ($E_{b}$) of various configurations
of CC-CC-ABNCNRs. NM and AFM denote the non-magnetic and anti-ferromagnetic
behaviors, respectively.} {M in the band gap column
denotes metallic behavior.}\label{tabl2}}
\end{table}

\begin{figure}[H]
\includegraphics{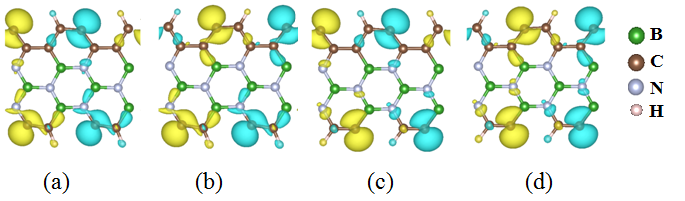}

\caption{{Spin density plots for (a) b$_{3}$, (b) b$_{4}$,
(c) b$_{5}$, and (d) b$_{6}$ configurations. Yellow and blue colors
represent spin-up, and spin-down, states, respectively}}

\label{spin_cc}
\end{figure}

\begin{figure}[H]
\includegraphics[scale=0.5]{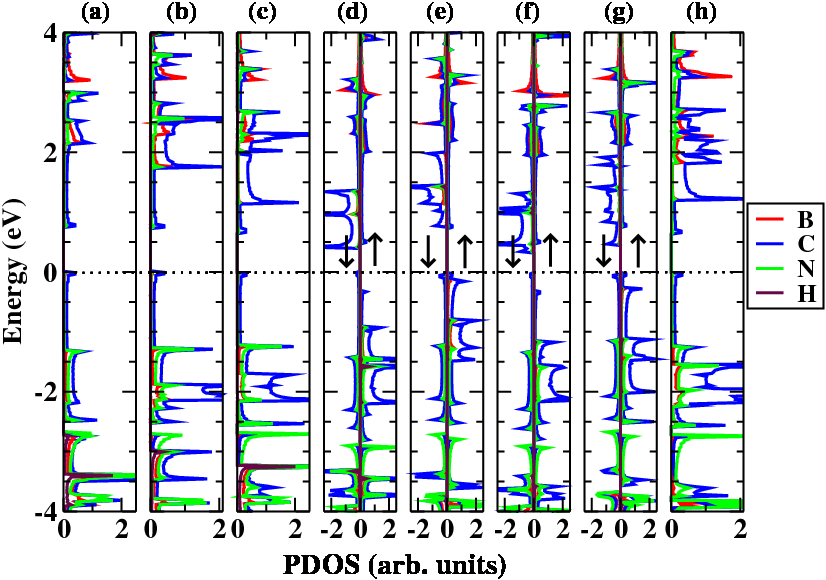}

\caption{Projected density of states of atoms B, C, N and H of all configurations
of 6-CC-CC-ABNCNR: (a) b\protect\textsubscript{0}, (b) b\protect\textsubscript{1},
(c) b\protect\textsubscript{2}, (d) b\protect\textsubscript{3},
(e) b\protect\textsubscript{4}, (f) b\protect\textsubscript{5},
(g) b\protect\textsubscript{6} and (h) b\protect\textsubscript{7}.
Up and down arrows indicate spin-up and spin-down states, respectively.
Fermi level has been set at 0 eV, and is represented by a black dashed
line.}

\label{pdos_cc}
\end{figure}

In Figs. \ref{Fig:band_cc} (i) and (j), (k) and (l), (m) and (n),
we present partial charge density profiles for configurations b\textsubscript{0},
b\textsubscript{5}, and b\textsubscript{7}, respectively. We note
that: (a) the VBM and CBM of configuration b\textsubscript{0} are
derived mainly from carbon atoms, (b) for b\textsubscript{5 }both
the VBM and CBM originate from the bare-edge atoms, with similar charge
distributions, and (c) in configuration b\textsubscript{7}, the VBM
derives its contributions from both the edges, while the CBM is localized
on the upper edge unpassivated C atoms. Similar to BN-BN-ABNCNRs,
CC-CC-ABNCNRs also exhibit spin polarization when one atom from the
upper edge, and another atom from the lower edge, are unpassivated.
However, unlike BN-BN-ABNCNRs, no metallic behavior is observed in
partially edge-hydrogenated CC-CC-ABNCNRs. PDOS of various configurations
of 6-CC-CC-ABNCNR is presented in Fig.\ref{pdos_cc}. From, the Fig.
\ref{pdos_cc}, it is clear that for both the nonmagnetic and the
magnetic configurations, VBM and CBM derive their main contributions
from the C atoms. In the spin-down channels of configurations b\textsubscript{3}-b\textsubscript{6},
the CBMs are mainly derived from the p\textsubscript{y} orbitals
of the unpassivated edge C atoms (Fig.1 of Supporting Information).
In {Fig. \ref{spin_cc} we present the spin-density
plots of configurations b\textsubscript{{3}}-b\textsubscript{{6}}.
From the Fig. \ref{spin_cc} and Table \ref{tabl2}, it is clear that
the spin-polarized configurations exhibit antiferromagnetic alignment
with zero net magnetic moments, because of equal number of spin-up
and spin-down states. It is also obvious from energetic considerations
that configuration b\textsubscript{{3}} is the most
stable magnetic configuration among all the magnetic configurations
of BN-BN-ABNCNRs (See Table \ref{tabl2}). }

{Table \ref{Bader_cc} presents the Bader charges on
the edge atoms of configurations b\textsubscript{{3}}
to b\textsubscript{{6 }}and compared with b\textsubscript{{7}}.
Except in the configuration b\textsubscript{{5}},
one hydrogen-passivated edge atom loses the electrons and another
one gains the electrons in configurations b\textsubscript{{3}},
b\textsubscript{{4}} and b\textsubscript{{6}}.
Since all the edge atoms are carbons, we did not find any significant
difference in the charges of hydrogen atoms.}

\begin{table}
\begin{centering}
\begin{tabular}{|c|c|c|c|c|c|c|c|c|}
\hline 
\multirow{2}{*}{{Configuration}} & \multirow{2}{*}{{LUE (C)}} & \multirow{2}{*}{{RUE(C)}} & \multirow{2}{*}{{LLE (C)}} & \multirow{2}{*}{{RLE (C)}} & \multirow{2}{*}{{H-LUE}} & \multirow{2}{*}{{H-RUE}} & \multirow{2}{*}{{H-LLE}} & \multirow{2}{*}{{H-RLE}}\tabularnewline
 &  &  &  &  &  &  &  & \tabularnewline
\hline 
{b\textsubscript{3}} & {3.89(-0.24)} & {4.18(-0.05)} & {3.39(-0.18)} & {4.95(+0.20)} & - & {0.930} & - & {0.910}\tabularnewline
\hline 
{b\textsubscript{4}} & {4.06(-0.07)} & {4.27(+0.04)} & {3.40(-0.17)} & {4.95(+0.20)} & {0.979} & - & - & {0.921}\tabularnewline
\hline 
{b\textsubscript{5}} & {4.00(-0.13)} & {4.11(-0.12)} & {3.80(+0.23)} & {4.53(-0.22)} & - & {0.998} & - & {0.954}\tabularnewline
\hline 
{b\textsubscript{6}} & {4.11(-0.02)} & {4.18(-0.05)} & {{} 3.87(+0.30)} & {4.52(-0.23)} & {0.950} & - & {0.975} & -\tabularnewline
\hline 
{b\textsubscript{7}} & {4.13 } & {4.23 } & {3.57} & {{} 4.75} & - & - & - & -\tabularnewline
\hline 
\end{tabular}
\par\end{centering}
\caption{{Bader charges on the edge atoms of configurations
of b\protect\textsubscript{{3}}-b\protect\textsubscript{{7}}.
The values given in the parentheses show the charge difference with
respect to the bare b\protect\textsubscript{{7}} configuration,
with positive/negative values denoting the electron gain/loss, respectively.
Rest of the notations have the same meaning as in the caption of Table
\ref{bader_bn}. }\label{Bader_cc}}
\end{table}

\subsection{Electronic properties of CC-BN-BNCNRS}

\begin{figure}[H]
\begin{centering}
\includegraphics[scale=0.75]{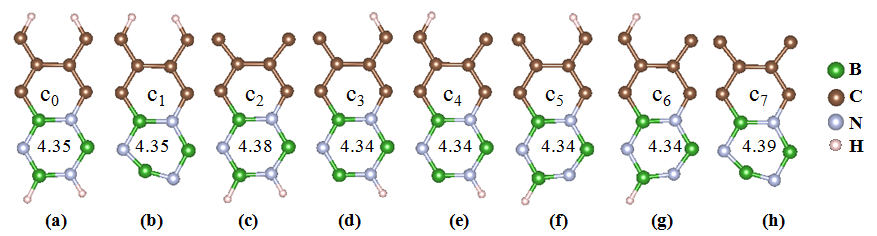}
\par\end{centering}
\caption{Geometric structures of eight configurations of CC-BN-ABNCNRs (Unit
cells): (a) C\protect\textsuperscript{H}C\protect\textsuperscript{H}-B\protect\textsuperscript{H}N\protect\textsuperscript{H},
(b) C\protect\textsuperscript{H}C\protect\textsuperscript{H}-BN,
(c) CC-B\protect\textsuperscript{H}N\protect\textsuperscript{H},
(d) CC\protect\textsuperscript{H}-BN\protect\textsuperscript{H},
(e) C\protect\textsuperscript{H}C-BN\protect\textsuperscript{H},
(f) CC\protect\textsuperscript{H}-B\protect\textsuperscript{H}N,
(g) C\protect\textsuperscript{H}C-B\protect\textsuperscript{H}N,
and (h) CC-BN. For the sake of brevity, we represent these configurations
by notations c\protect\textsubscript{0}through c\protect\textsubscript{7 }as
shown in the figure. {Lattice constants for all the
configurations are displayed in \AA{} units.}}

\label{Fig:str_cn}
\end{figure}

Like previously discussed ribbons, CC-BN-ABNCNRs are also divided
into eight configurations based on the edge atom passivation scheme:
C\textsuperscript{H}C\textsuperscript{H}-B\textsuperscript{H}N\textsuperscript{H}-ABNCNR
(c\textsubscript{0}), C\textsuperscript{H}C\textsuperscript{H}-BN-ABNCNR
(c\textsubscript{1}), CC-B\textsuperscript{H}N\textsuperscript{H}-ABNCNR
(c\textsubscript{2}), CC\textsuperscript{H}-BN\textsuperscript{H}-ABNCNR
(c\textsubscript{3}), C\textsuperscript{H}C-BN\textsuperscript{H}-ABNCNR
(c\textsubscript{4}), CC\textsuperscript{H}-B\textsuperscript{H}N-ABNCNR
(c\textsubscript{5}), C\textsuperscript{H}C-B\textsuperscript{H}N-ABNCNR
(c\textsubscript{6}), CC-BN-ABNCNR (c\textsubscript{7}). All the
geometrical structures of partially hydrogen-passivated CC-BN-ABNCNRs
are presented in Fig. \ref{Fig:str_cn}, from which it is clear that
configurations c\textsubscript{0} and c\textsubscript{7} represent
fully hydrogen-passivated, and bare CC-BN ABNCNRs, respectively. {These
ABNCNRs have lattice constants in the range 4.34 to 4.39 \AA{}
(See Fig. \ref{Fig:str_cn}).}

\begin{figure}[H]
\begin{centering}
\includegraphics[scale=0.8]{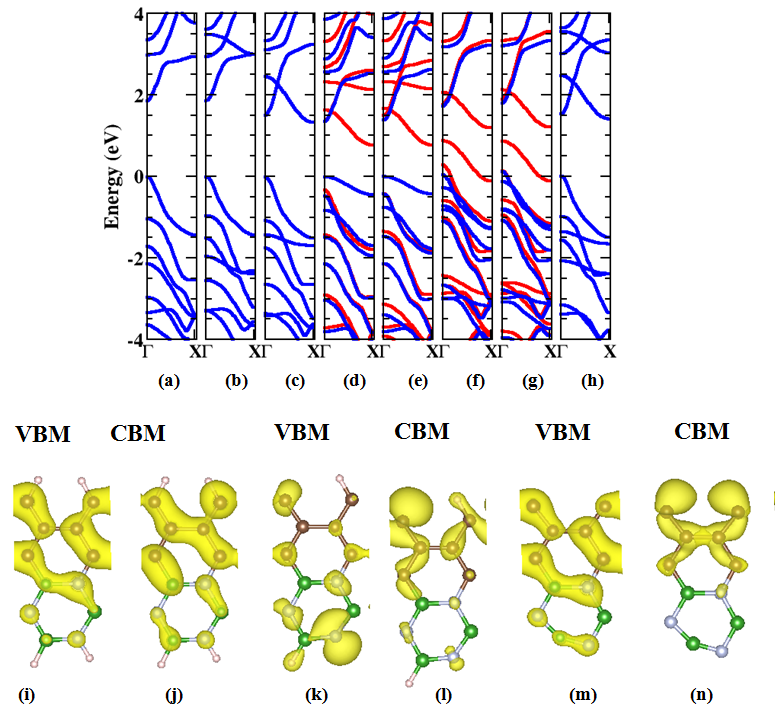}
\par\end{centering}
\caption{Band structures of (a) c\protect\textsubscript{0}, (b) c\protect\textsubscript{1},
(c) c\protect\textsubscript{2}, (d) c\protect\textsubscript{3},
(e) c\protect\textsubscript{4}, (f) c\protect\textsubscript{5},
(g) c\protect\textsubscript{6}, and (h) c\protect\textsubscript{7}.
Partial charge density profiles: (i) and (j) VBM and CBM of c\protect\textsubscript{0},
(k) and (l) VBM and CBM of c\protect\textsubscript{5}, (m) and (n)
VBM and CBM of c\protect\textsubscript{7}. Blue and red curves denote
bands corresponding to opposite spin orientations. Colors of atoms
are consistent with Fig. \ref{Fig:str_cn}}

\label{Fig:band_cn}
\end{figure}

Electronic structures of various CC-BN-ABNCNRs configurations are
presented in Fig. \ref{Fig:band_cn}, while Table \ref{table3} contains
their band gaps, magnetic moments, and cohesive energies. Figs. \ref{Fig:band_cn}
(a)-(c) and (h) display band structures for configurations c\textsubscript{0},
c\textsubscript{1}, c\textsubscript{2}, and c\textsubscript{7},
making it obvious that they are all non-magnetic semiconductors, out
of which c\textsubscript{0} and c\textsubscript{1} have direct band
gaps. We note that configurations c\textsubscript{2} and c\textsubscript{7}
have indirect band gaps, while in comparison, for BN-BN-, and CC-CC-ABNCNRs,
band gaps of all non-magnetic semiconductors are direct in nature.
From Figs. \ref{Fig:band_cn} (d)-(g), and Table \ref{table3}, based
upon split bands, and different band gaps for the two spin orientations,
we conclude that the nanoribbons with configurations c\textsubscript{3}-c\textsubscript{6},
exhibit spin-polarized behavior. In configurations $c_{3}$ and $c_{4}$,
both the band gaps have semiconducting behavior, while in configurations
c\textsubscript{5 }and c\textsubscript{6} the valence and the conduction
bands cross Fermi-level, thereby rendering them metallic. But, the
metallicity in c\textsubscript{5} and c\textsubscript{6} is magnetic
in nature because of the spin-polarized character of their band structure,
and finite magnetic moments. Therefore, similar to the case of BN-BN-ABNCNRs,
we can also tune CC-BN-ABNCNRs into semiconductors, magnetic semiconductors,
or magnetic metals by manipulating scheme of partial edge passivation. 

\begin{table}[H]
\begin{centering}
\begin{tabular}{|c|c|c|c|c|c|c|c|c|}
\hline 
\multirow{2}{*}{Configuration} & \multicolumn{2}{c|}{Band gap (eV) } & \multirow{2}{*}{{$MM$ ($\mu_{B}$)}} & \multirow{2}{*}{{Type}} & \multirow{2}{*}{{$\Delta E$ (eV)}} & \multirow{2}{*}{{$\delta G$ (eV)}} & \multirow{2}{*}{{$E_{f}$ (eV/\AA )}} & \multirow{2}{*}{{$E_{b}$ (eV)}}\tabularnewline
\cline{2-3} 
 & Up  & Down &  &  &  &  &  & \tabularnewline
\hline 
c\textsubscript{0} & 1.85 & 1.85 & {0} & {NM} & {-} & {0.093 } & {0.172} & {-5.075}\tabularnewline
\hline 
c\textsubscript{1} & 1.86 & 1.86 & {0} & {NM} & {-} & {0.307} & {0.493} & {-5.400}\tabularnewline
\hline 
c\textsubscript{2} & 1.32 & 1.32 & {0} & {NM} & {-} & {0.398} & {0.490} & {-4.908}\tabularnewline
\hline 
c\textsubscript{3} & 1.37 & 1.10 & {1.08} & {FM} & {0.674 } & {0.459} & {0.740} & {-4.484}\tabularnewline
\hline 
c\textsubscript{4} & 1.34 & 1.18 & {0} & {AFM} & {0.197} & {0.456} & {0.735} & {-4.503}\tabularnewline
\hline 
c\textsubscript{5} & M & M & {1.23} & {FM} & {0.321} & {0.496} & {0.800} & {-4.222}\tabularnewline
\hline 
c\textsubscript{6} & M & M & {0} & {AFM} & {0.357} & {0.494} & {0.798} & {-4.232}\tabularnewline
\hline 
c\textsubscript{7} & 1.49 & 1.49 & {0} & {NM} & {-} & {0.699} & {0.956} & {-}\tabularnewline
\hline 
\end{tabular}
\par\end{centering}
\caption{Band gaps for the two spin orientations, magnetic moments per unit
cell ($MM)$, {type of magnetic behavior, energy difference
between magnetic and nonmagnetic states ($\Delta E$), Gibbs free
energy of formation per atom ($\delta G$), edge formation energy
per unit length ($E_{f}$), and binding energy per H atom ($E_{b}$)
of various configurations of CC-BN-ABNCNRs.} {NM, FM
and AFM represent nonmagnetic, ferromagnetic, and anti-ferromagnetic
behaviors, respectively. M in the band gap column implies metallic
behavior. }\label{table3}}
\end{table}

Figs. \ref{Fig:band_cn} (i) and (j), (k) and (l), and (m) and (n)
present partial charge density profiles for configurations c\textsubscript{0},
c\textsubscript{5,}and c\textsubscript{7}, respectively. The VBM
and CBM of c\textsubscript{0} have dominant presence mainly on carbon
atoms, with similar charge distributions. In c\textsubscript{5},
the VBM has dominant presence on the lower BN edge, while CBM originates
mainly from the upper edge C atoms (see Figs. \ref{Fig:band_cn} (k)
and (l)). In case of configuration c\textsubscript{7}, the charge
density corresponding to the VBM is delocalized significantly over
the entire width, while for the CBM it is mainly localized on upper
edge carbon atoms. 

We note that for CC-BN-ABNCNRs, metallic behavior is observed only
when C atom from one edge, and N atom from another edge are left bare.
As far as spin polarization is concerned, similar to the other two
classes of ABNCNRs previously discussed, these nanoribbons exhibit
it only when one atom on each edge is passivated.

\begin{figure}[H]
\includegraphics[scale=0.5]{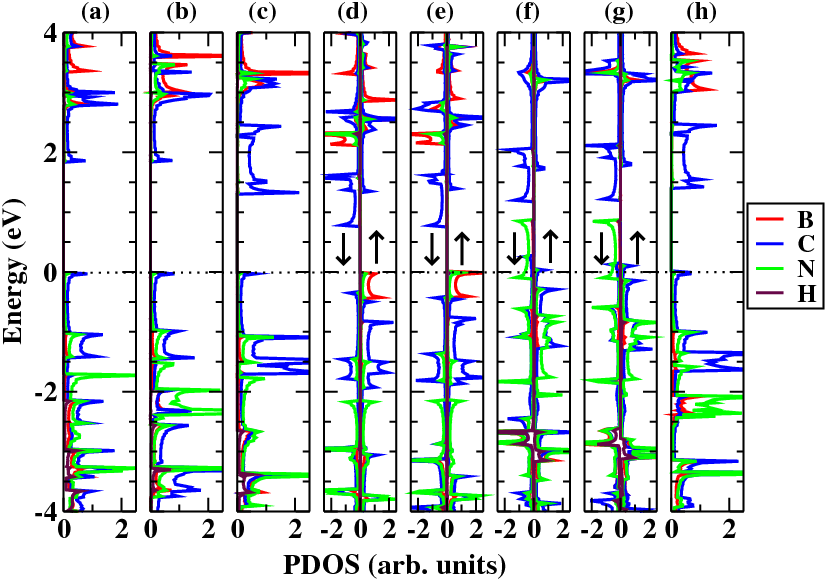}

\caption{Projected density of states of atoms B, C, N and H of all configurations
of 6-CC-BN-ABNCNR: (a) c\protect\textsubscript{0}, (b) c\protect\textsubscript{1},
(c) c\protect\textsubscript{2}, (d) c\protect\textsubscript{3},
(e) c\protect\textsubscript{4}, (f) c\protect\textsubscript{5},
(g) c\protect\textsubscript{6} and (h) c\protect\textsubscript{7}.
Up and down arrows indicate spin-up and spin-down, states, respectively.
Fermi level has been set at 0 eV, and is denoted causing them to become
metallic, by a black dashed line.}

\label{pdos_cn}
\end{figure}

\begin{figure}[H]
\includegraphics{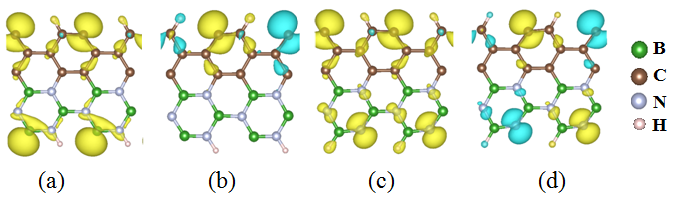}

\caption{{Spin density plots for (a) c$_{3}$, (b) c$_{4}$,
(c) c$_{5}$, and (d) c$_{6}$ configurations. Yellow and blue colors
represent spin-up, and spin-down, states, respectively}}

\label{spin_cn}
\end{figure}

Fig.\ref{pdos_cn} presents PDOS of all the configurations of 6-CC-BN-ABNCNR.
Similar to the BN-BN-ABNCNRS, VBM and CBM of nonmagnetic CC-BN-ABNCNR
configurations are derived from the C atoms. In the spin-up channel
configurations of c\textsubscript{3} and c\textsubscript{4}, VBMs
are mainly contributed by B atoms, with small contributions also from
the N atoms. In the spin-down channel, CBMs are derived essentially
from the C atoms. In the c\textsubscript{5} and c\textsubscript{6}
configurations, additional bands arise due to the following contributions:
(a) VBMs originate from the C atoms in the spin-up channel, and (b)
N atoms contribute to the VBM in the spin-down channel. These additional
energy bands around the Fermi level turn them into metals. We also
present PDOS of $p$ orbitals of the bare edge atoms of spin-polarized
configurations (c\textsubscript{3}-c\textsubscript{6}) in Fig. 3
of the Supporting Information. It is obvious that the metallic behavior
in c\textsubscript{5} and c\textsubscript{6} configurations is originating
from the anti-bonding electrons of p\textsubscript{y} orbitals. {Spin
density plots of configurations c\textsubscript{3}-c\textsubscript{6}
are presented in the Fig. \ref{spin_cn}. From Table \ref{table3}
and Fig. \ref{spin_cn} it is clear that the electrons from the unpassivated
edge atoms contribute to the magnetic moments, with c\textsubscript{3}
and c\textsubscript{5} exhibiting ferromagnetic behavior, and c\textsubscript{{4}}
and c\textsubscript{{6}} showing antiferromagnetic
alignment. Energetically speaking, the most stable configuration among
all the magnetic configurations of CC-BN-ABNCNRs is c\textsubscript{3}.}

{Bader charges on the edge atoms after the partial
passivation are presented in the table \ref{Bader_cn} for configurations
c\textsubscript{{3}}-c\textsubscript{{6}},
and are compared with those on the bare CC-BN configuration (c7).
In configuration c\textsubscript{{3}}, except RUE,
all other edge atoms gain electronic charge. In configurations c\textsubscript{{5}}
and c\textsubscript{{6}}, both LUE and LLE atoms gain
electrons. Interestingly, in the configuration c\textsubscript{{4}}
two edge atoms gain the electronic charge, while another two lose
it. In configurations c\textsubscript{{3}} and c\textsubscript{{4}},
N atoms passivated by H atoms have low charges, while in the configuration
c\textsubscript{{5}} and c\textsubscript{{6}},
H-RLE and H-LLE are found to have high electronic charges.}
\begin{table}
\begin{centering}
\begin{tabular}{|c|c|c|c|c|c|c|c|c|}
\hline 
\multirow{2}{*}{{Configuration}} & \multirow{2}{*}{{LUE (C)}} & \multirow{2}{*}{{RUE(C)}} & \multirow{2}{*}{{LLE (B)}} & \multirow{2}{*}{{RLE (N)}} & \multirow{2}{*}{{H-LUE}} & \multirow{2}{*}{{H-RUE}} & \multirow{2}{*}{{H-LLE}} & \multirow{2}{*}{{H-RLE}}\tabularnewline
 &  &  &  &  &  &  &  & \tabularnewline
\hline 
{c\textsubscript{3}} & {4.20(+0.15)} & {3.94 (-0.23)} & {1.60 (+0.37)} & {6.74 (+0.19)} & {-} & {0.976} & {-} & {0.625}\tabularnewline
\hline 
{c\textsubscript{4}} & {3.94(-0.11)} & {4.15(-0.02)} & {1.61(+0.38)} & {6.72(+0.17)} & {0.971} & {-} & {-} & {0.629}\tabularnewline
\hline 
{c\textsubscript{5}} & {4.27(+0.22)} & {3.93(-0.24)} & {1.22(-0.01)} & {6.39(-0.16)} & {-} & {0.973} & {-} & {1.601}\tabularnewline
\hline 
{c\textsubscript{6}} & {4.01(-0.04)} & {3.98(-0.09)} & {1.27(+0.04)} & {6.35(-0.20)} & {0.946} & {-} & {1.600} & {-}\tabularnewline
\hline 
{c\textsubscript{7}} & {4.05} & {4.17} & {1.23} & {6.55} & {-} & {-} & {-} & {-}\tabularnewline
\hline 
\end{tabular}
\par\end{centering}
\caption{{Charges on the edge atoms of configurations of c\protect\textsubscript{{3}}-c\protect\textsubscript{{7}}.
Positive and negative values denote the electron gain and the electron
loss, respectively. The values given in the parentheses show the charge
difference, compared to those on the corresponding atoms of the bare
configuration c\protect\textsubscript{{7}}. Rest of
the information is same as in the caption of Table \ref{bader_bn}.
}\label{Bader_cn}}
\end{table}

\subsection{Relative Stability of ABNCNRs}

{In this section we discuss the relative stability
of various configurations of ABNCNRs, based upon their zero temperature
Gibbs free energies of formation ($\delta$G) presented in the Tables
\ref{table1}, \ref{tabl2}, and \ref{table3}. $\delta G$ was computed
using the following formula which has been used by other authors for
structures with a variety of chemical compositions\cite{silicane,kan_2008,ohod,kan_jacs}. }

{
\[
\delta G=\left(E_{c}-x_{B}\mu_{B}-x_{N}\mu_{N}-x_{C}\mu_{C}-x_{H}\mu_{H}\right),
\]
}

{where $\mu_{C}$, $\mu_{B}$, $\mu_{H}$, and $\mu_{N}$
are the chemical potentials of atoms C, B, H, and N, respectively;
$x_{C}$, $x_{B}$, $x_{H}$, and $x_{N}$ are the molar fractions
of C, B, H, and N atoms, respectively, satisfying the rule $x_{C}+x_{B}+x_{H}+x_{N}=1$,
and $\mu_{C}$ is defined as the cohesive energy per atom in the infinite
monolayer graphene sheet. We can write the equation above as 
\[
\delta G=\left(E_{c}-x_{BN}\mu_{BN}-x_{C}\mu_{C}-x_{H}\mu_{H}\right),
\]
}{where $E_{c}$ is the cohesive energy per atom
of the ribbon in question; $\mu_{BN}$ denotes the
chemical potential of the BN dimer calculated as cohesive energy per
unit cell of the 2D monolayer of BN, and $x_{BN}=x_{B}+x_{N}$ represents
the molar fraction of BN units. $\mu_{H}$ is computed as the binding
energy per atom of H\textsubscript{2} molecule\cite{kan_2008}, and
to account for the thermodynamic equilibrium of $\mu_{B}$ and $\mu_{N}$
, we used the constraint : $\mu_{BN}=\mu_{B}+\mu_{N}$.\cite{akmanna2,bcn1}
Cohesive energy per atom of the ribbon $E_{c}$ is defined as
\[
E_{c}=\left(E_{T}-n_{B}E_{B}-n_{C}E_{C}-n_{N}E_{N}-n_{H}E_{H}\right)/n_{T},
\]
}

{above $E_{C}$, $E_{B}$, $E_{H}$, and $E_{N}$ are
the total energies of isolated atoms C, B, H, and N, respectively;
$n_{C}$, $n_{B}$, $n_{H}$, and $n_{N}$ are the numbers of C, B,
H, and N atoms, respectively, in the unit cell, while $n_{T}=n_{C}+n_{B}+n_{H}+n_{N}$,
is the total number of atoms per cell.}

{Generally a lower Gibbs free energy of formation implies
a higher stability of the system. As we have already discussed in
earlier sections that fully edge-hydrogenated ABNCNRs (a\textsubscript{0},
b\textsubscript{0}, and c\textsubscript{0}) have the relative stability
in the order c\textsubscript{0 }$>$ b\textsubscript{0 }$>$a\textsubscript{0},
while bare ABNCNRs exhibit the order c\textsubscript{7 }$>$ a\textsubscript{7 }$>$b\textsubscript{7}. }

{We also compared partially edge-hydrogenated ABNCNRs,
and found that within a class of nanoribbons, the following stability
order is observed: (i) a\textsubscript{1}$>$ a\textsubscript{2}
$>$ a\textsubscript{3} $>$ a\textsubscript{4}$>$ a\textsubscript{6}
$>$ a\textsubscript{5}, (ii) b\textsubscript{1}$>$ b\textsubscript{2}
$>$ b\textsubscript{4} $>$ b\textsubscript{6}$>$ b\textsubscript{3}
$>$ b\textsubscript{5}, and (iii) c\textsubscript{1}$>$ c\textsubscript{2}
$>$ c\textsubscript{4} $>$ c\textsubscript{3}$>$ c\textsubscript{6}
$>$ c\textsubscript{5}. The common feature among these orders is
that the most stable of the partially passivated ribbons in a given
class is the ribbon whose both upper edge atoms are saturated. The
least stable configurations also have a common feature in that one
upper, and another lower edge atom are saturated, which are diagonally
across, and topologically similar, across the three classes of nanoribbons
(see Figs. \ref{Fig:str_bn}, \ref{Fig:str_cc}, and \ref{Fig:str_cn}). }

\begin{figure}[H]
\includegraphics[scale=0.5]{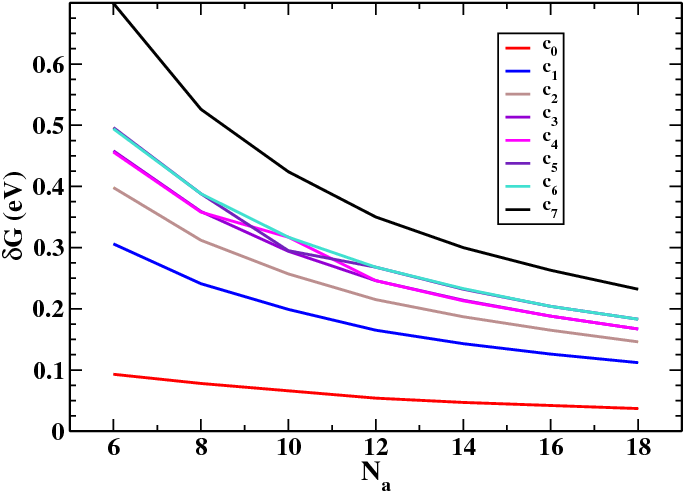}

\caption{{Gibbs free energy of formation per atom of all the
configurations of CC-BN-ABNCNRs, as a function of ribbon width.}}

\label{Fig:gibbs-1}
\end{figure}

{We also present the behavior of the Gibbs free energies
of the most stable configurations of CC-BN-ABNCNRs, as a function
of their width in the Fig. \ref{Fig:gibbs-1}. We find that the Gibbs
free energies of all the configurations decrease with the increasing
width, implying that the ribbons with larger widths are more stable
compared to the narrower ones, similar to ABNNRs\cite{half_bnnr}.}

{We also calculated edge formation energy per unit
length ($E_{f}$)\cite{Nalaal1} of ABNCNRs by using the formula }

{
\[
E_{f}=\left(E_{T}-n_{BN}E_{BN}-n_{C}E_{C}-0.5n_{H}E_{H_{2}}\right)/2L,
\]
}

{where $E_{T}$ is the total energy/cell of the
nanoribbon, $E_{BN}$ is the energy per BN dimer of BN sheet, $E_{H_{2}}$
is the total energy of H\textsubscript{2} molecule, $n_{BN}$ is
the total number of BN dimers, and L is the length
of the edge (in \AA{} ) of the given nanoribbon. The binding energy
per H atom ($E_{b}$)\cite{bnc_zig} for the given ribbon was computed
as}

{
\[
E_{b}=\left(E_{H-ABNCNR}-E_{ABNCNR}-n_{H}E_{H}\right)/n_{H},
\]
}

\begin{figure}[H]
\includegraphics[scale=0.5]{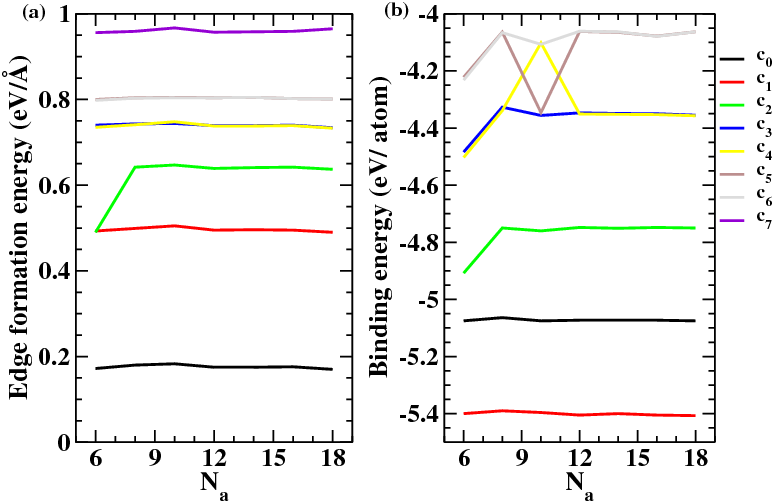}\caption{(a) {Edge formation energy ($E_{f}$), and (b) the
binding energy per H atom ($E_{b}$) of CC-BN-ABNCNRS as a function
of width} }

\label{form_bind}
\end{figure}

{where $E$$_{H-ABNCNR}$\textsubscript{} is the total
energy of partially or fully hydrogenated ABNCNR and $E_{ABNCNR}$\textsubscript{\textsubscript{}}
is the total energy of the corresponding bare ABNCNR. $E_{f}$ and
$E_{b}$ are presented in Tables \ref{table1}, \ref{tabl2}, and
\ref{table3} for N\textsubscript{{a}}=6, for various
ABNCNRs. For all the configurations of CC-BN-ABNCNRs, we have presented
the width dependence of $E_{f}$ and $E_{b}$ in Figs. \ref{form_bind}
(a) and (b), respectively. From the tables (See Tables \ref{table1},
\ref{tabl2}, and \ref{table3}), it is clear that $E_{f}$ exhibits
similar trends as $\delta G$ for Na=6, implying that the fully hydorgenated
ABNCNRs are more favorable towards edge formation than the ones which
are partially passivated, or completely bare. Considering the $E_{b}$
values for Na=6 (See Tables \ref{table1}, \ref{tabl2}, and \ref{table3}),
we found that the fully and the partially edge hydrogenated ABNCNRs
have negative binding energies in the range of -5.40 eV/atom to -3.81
eV/atom. This indicates that H atoms bind strongly to the edge atoms,
preventing their dissociation from the edges\cite{bnc_zig}. For the
partially passivated nanoribbons (BN-BN, CC-CC, and CC-BN ABNCNRs),
configurations a\textsubscript{{1}}, b\textsubscript{{1}}
and c\textsubscript{{1}} have high binding energies,
while a\textsubscript{{5}}, b\textsubscript{{5}},
and c\textsubscript{{5}} have relatively lower binding
energies, implying that it is easier to hydrogenate upper edge passivated
nanoribbons, as compared to other partially passivated ABNCNRs. From
Figs. \ref{form_bind} (a) and (b), it is clear that both the energies
$E_{f}$ and $E_{b}$ have weak width dependence.}

\subsection{Width Dependence of Properties}

\begin{figure}
\includegraphics[scale=0.5]{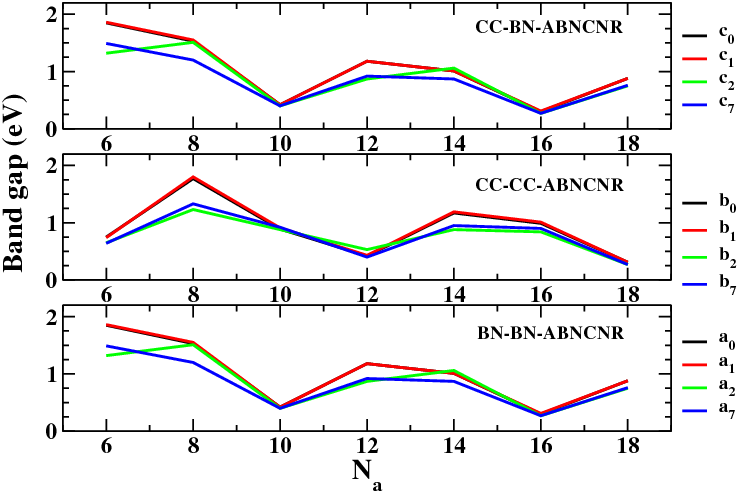}

\caption{Band gaps of nonmagnetic CC-BN-, CC-CC-, and BN-BN-ABNCNRs, as functions
of ribbon widths.}

\label{Fig:width}
\end{figure}

Quantitative calculations presented in the previous sections were
performed for nanoribbons with the width $N_{a}=6$. Therefore, the
question arises: Will the qualitative features exhibited by ribbons
of this width also hold for wider ribbons? To ascertain that we performed
calculations for the ribbons with the widths in the range 6$\leq N_{a}\leq\text{18}$,
and found that wider CC-BN-ABNCNRs, BN-BN-ABNCNRs, and CC-CC-ABNCNRs
exhibit behavior similar to that observed for width $N_{a}=6$. Fig.
\ref{Fig:width} presents the band gaps of nonmagnetic ABNCNRs, as
a function of their width, and Table \ref{Tab:width} presents the
band gaps of spin-polarized CC-BN-ABNCNRS (c\textsubscript{3}, c\textsubscript{4},
c\textsubscript{5} and c\textsubscript{6}). From Fig. \ref{Fig:width},
we conclude that overall band gaps of nonmagnetic ABNCNRs decrease
with the increasing width, in an oscillatory behavior, similar to
the cases of AGNRs\cite{AGNR_Eg_osc} and ABNNRs\cite{ABNNR_Eg_osc}.
Band structures of all configurations of BN-BN-, CC-CC-, and CC-BN
ABNCNRs for width $N_{a}=18$ are presented in Figs. 5-7 of the Supporting
Information. From Table.\ref{Tab:width}, it is clear that similar
to the nonmagnetic ABNCNRs, spin-polarized band gaps (up and down
channels) of c$_{3}$ and c$_{4}$ configurations also decrease with
the increasing the width, while c\textsubscript{5} and c\textsubscript{6}
configurations are metallic for all the widths. Spin-polarized band
gaps of CC-CC-ABNCNRs and BN-BN-ABNCNRs are presented in Tables 1
and 2, respectively, of the Supporting Information. Similar to CC-BN-ABNCNRs,
the band gaps of spin-polarized configurations of BN-BN (a$_{3}$)
and CC-CC-ABNCNRs(b$_{1}$-b$_{6}$) decrease with the increasing
width. And, configurations a\textsubscript{4}-a\textsubscript{6}
exhibit metallic behavior irrespective of their widths. Thus, we believe
that the qualitative features observed for ribbons of width $N_{a}=6$,
hold also for much wider ribbons. 

\begin{table}
\begin{tabular}{|c|c|c|c|c|c|c|c|c|}
\hline 
\multirow{2}{*}{$N_{a}$} & \multicolumn{2}{c}{} & \multicolumn{2}{c}{Band gaps (eV)} & \multicolumn{2}{c}{} & \multicolumn{2}{c|}{}\tabularnewline
\cline{2-9} 
 & \multicolumn{2}{c|}{c\textsubscript{3}} & \multicolumn{2}{c|}{c\textsubscript{4}} & \multicolumn{2}{c|}{c\textsubscript{5}} & \multicolumn{2}{c|}{c\textsubscript{6}}\tabularnewline
\cline{2-9} 
 & Up & Dn & Up & Dn & Up & Dn & Up & Dn\tabularnewline
\hline 
8 & 1.38 & 0.84 & 1.40 & 0.95 & M & M & M & M\tabularnewline
\hline 
10 & 0.45 & 0.23 & 0.38 & 0.48 & M & M & M & M\tabularnewline
\hline 
12 & 1.08 & 0.74 & 1.09 & 0.80 & M & M & M & M\tabularnewline
\hline 
14 & 1.03 & 0.61 & 1.01 & 1.11 & M & M & M & M\tabularnewline
\hline 
16 & 0.33 & 0.18 & 0.27 & 0.37 & M & M & M & M\tabularnewline
\hline 
18 & 0.88 & 0.57 & 0.86 & 0.65 & M & M & M & M\tabularnewline
\hline 
\end{tabular}

\caption{Band gaps of spin-polarized CC-BN-ABNCNRs as a function of ribbon
width ($N_{a}$). M denotes that the corresponding ribbon is metallic.
\label{Tab:width}}
\end{table}

\section{Conclusions}

\label{sec:conclusions}

We performed extensive electronic structure calculations on fully
and partially hydrogen-passivated ABNCNRs, using the first-principles
based DFT approach, and studied three types of configurations, namely,
CC-CC-, BN-BN- and CC-BN-, depending upon the atoms present on their
edges. When compared to fully hydrogen-passivated ABNCNRs, partial
edge passivation causes the formation of additional energy bands near
Fermi-levels of these nanoribbons, leading to different electronic
properties. We found that fully hydrogen-passivated, and completely
bare ABNCNRs, are non-magnetic semiconductors. However, as far as
partially passivated ABNCNRs are concerned, they were found to exhibit
a wide range of electronic behavior such as normal metal, magnetic
metal, normal semiconductor, and spin-polarized semiconductor, depending
on their edge termination, and hydrogenation schemes. We also found
that the ABNCNRs with larger widths are more stable compared to the
narrower ones. Furthermore, a few their configurations exhibited metallic
behavior, for all the widths considered. Thus, in principle, partial
hydrogenation of edges allows the tuning of the band structures of
ABNCNRs, with possible applications in the fields of spintronics,
and optoelectronic devices.
\section{acknowledgments}
NA and NM gratefully acknowledge the support from Monash HPC, National
Computing Infrastructure of Australia, and the Pawsey Supercomputing
facility. This research was partially supported by the Australian
Research Council Centre of Excellence in Future Low-Energy Electronics
Technologies (project number CE170100039) and funded by the Australian
Government. AS acknowledges the financial support from Department
of Science and Technology, Government of India, under project no.
SB/S2/CMP-066/2013.
%\end{acknowledgments}

\bibliographystyle{rsc}
\addcontentsline{toc}{section}{\refname}\bibliography{abncnr}
\newpage{}
\end{document}

\pagestyle{fancy}
\thispagestyle{plain}
\fancypagestyle{plain}{

%%%HEADER%%%
\fancyhead[C]{\includegraphics[width=18.5cm]{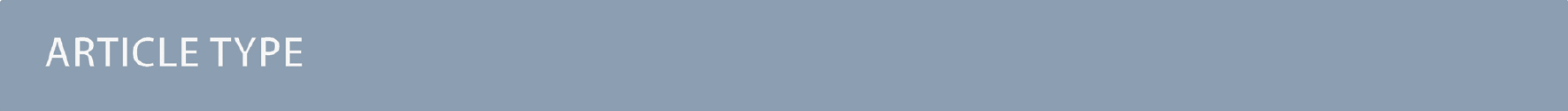}}
\fancyhead[L]{\hspace{0cm}\vspace{1.5cm}\includegraphics[height=30pt]{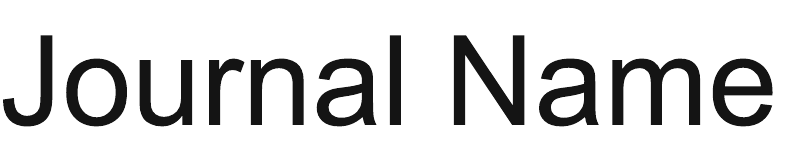}}
\fancyhead[R]{\hspace{0cm}\vspace{1.7cm}\includegraphics[height=55pt]{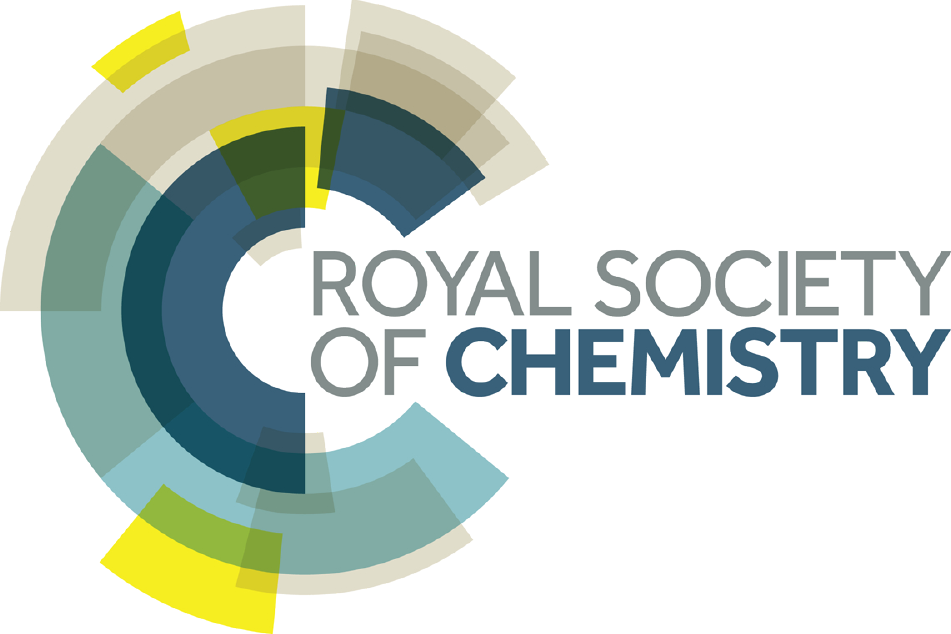}}
\renewcommand{\headrulewidth}{0pt}
}
%%%END OF HEADER%%%

%%%PAGE SETUP - Please do not change any commands within this section%%%
\makeFNbottom
\makeatletter
\renewcommand\LARGE{\@setfontsize\LARGE{15pt}{17}}
\renewcommand\Large{\@setfontsize\Large{12pt}{14}}
\renewcommand\large{\@setfontsize\large{10pt}{12}}
\renewcommand\footnotesize{\@setfontsize\footnotesize{7pt}{10}}
\renewcommand\scriptsize{\@setfontsize\scriptsize{7pt}{7}}
\makeatother

\renewcommand{\thefootnote}{\fnsymbol{footnote}}
\renewcommand\footnoterule{\vspace*{1pt}% 
\color{cream}\hrule width 3.5in height 0.4pt \color{black} \vspace*{5pt}} 
\setcounter{secnumdepth}{5}

\makeatletter 
\renewcommand\@biblabel[1]{#1}            
\renewcommand\@makefntext[1]% 
{\noindent\makebox[0pt][r]{\@thefnmark\,}#1}
\makeatother 
\renewcommand{\figurename}{\small{Fig.}~}
\sectionfont{\sffamily\Large}
\subsectionfont{\normalsize}
\subsubsectionfont{\bf}
\setstretch{1.125} %In particular, please do not alter this line.
\setlength{\skip\footins}{0.8cm}
\setlength{\footnotesep}{0.25cm}
\setlength{\jot}{10pt}
\titlespacing*{\section}{0pt}{4pt}{4pt}
\titlespacing*{\subsection}{0pt}{15pt}{1pt}
%%%END OF PAGE SETUP%%%

%%%FOOTER%%%
\fancyfoot{}
\fancyfoot[LO,RE]{\vspace{-7.1pt}\includegraphics[height=9pt]{LF}}
\fancyfoot[CO]{\vspace{-7.1pt}\hspace{13.2cm}\includegraphics{RF}}
\fancyfoot[CE]{\vspace{-7.2pt}\hspace{-14.2cm}\includegraphics{RF}}
\fancyfoot[RO]{\footnotesize{\sffamily{1--\pageref{LastPage} ~\textbar  \hspace{2pt}\thepage}}}
\fancyfoot[LE]{\footnotesize{\sffamily{\thepage~\textbar\hspace{3.45cm} 1--\pageref{LastPage}}}}
\fancyhead{}
\renewcommand{\headrulewidth}{0pt} 
\renewcommand{\footrulewidth}{0pt}
\setlength{\arrayrulewidth}{1pt}
\setlength{\columnsep}{6.5mm}
\setlength\bibsep{1pt}
%%%END OF FOOTER%%%

%%%FIGURE SETUP - please do not change any commands within this section%%%
\makeatletter 
\newlength{\figrulesep} 
\setlength{\figrulesep}{0.5\textfloatsep} 

\newcommand{\topfigrule}{\vspace*{-1pt}% 
\noindent{\color{cream}\rule[-\figrulesep]{\columnwidth}{1.5pt}} }

\newcommand{\botfigrule}{\vspace*{-2pt}% 
\noindent{\color{cream}\rule[\figrulesep]{\columnwidth}{1.5pt}} }

\newcommand{\dblfigrule}{\vspace*{-1pt}% 
\noindent{\color{cream}\rule[-\figrulesep]{\textwidth}{1.5pt}} }

\makeatother
%%%END OF FIGURE SETUP%%%

%%%TITLE AND AUTHORS%%%
\twocolumn[
  \begin{@twocolumnfalse}
\vspace{3cm}
\sffamily
\begin{tabular}{m{4.5cm} p{13.5cm} }

\includegraphics{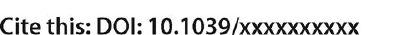} & \noindent\LARGE{\textbf{Supporting Information for Tunable electronic properties of partially edge-hydrogenated armchair
boron-nitrogen-carbon nanoribbons}} \\%Article title goes here instead of the text "This is the title"
 & \vspace{0.3cm} \\

 & \noindent\large{Naresh Alaal,\textit{$^{a,b,c}$} Nikhil Medhekar,\textit{$^{d}$} and Alok Shukla\textit{$^{b,c,\ddag}$}} \\%Author names go here instead of "Full name", etc.

\includegraphics{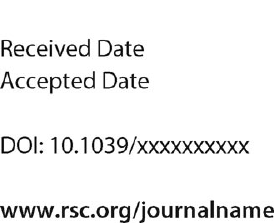} & \\

\end{tabular}

 \end{@twocolumnfalse} \vspace{0.6cm}

  ]
%%%END OF TITLE AND AUTHORS%%%

%%%FONT SETUP - please do not change any commands within this section
\renewcommand*\rmdefault{bch}\normalfont\upshape
\rmfamily
\section*{}
\vspace{-1cm}

%%%FOOTNOTES%%%

\footnotetext{\textit{$^{a}$~IITB-Monash Research Academy, CSE Building 2$^{\,nd}$ Floor,
IIT Bombay, Mumbai 400076, India}}
\footnotetext{\textit{$^{b}$~Department of Physics, Indian Institute of Technology Bombay, Powai, Mumbai 400076 India}}
\footnotetext{\textit{$^{c}$~Department of Physics, Bennett University,
Plot No 8-11, TechZone II, Greater Noida 201310 (UP), INDIA }}
\footnotetext{\textit{$^{d}$~Department of Materials Engineering, Monash University, Clayton,
Victoria 3800, Australia}}

%Please use \dag to cite the ESI in the main text of the article.
%If you article does not have ESI please remove the the \dag symbol from the title and the footnotetext below.
%\footnotetext{\dag~Electronic Supplementary Information (ESI) available: [details of any supplementary information available should be included here]. See DOI: 10.1039/b000000x/}
%additional addresses can be cited as above using the lower-case letters, c, d, e... If all authors are from the same address, no letter is required

\footnotetext{\ddag~E-mail: shukla@phy.iitb.ac.in}

%%%END OF FOOTNOTES%%%
\onecolumn
\begin{figure}[H]
\centering{}\includegraphics[scale=0.5]{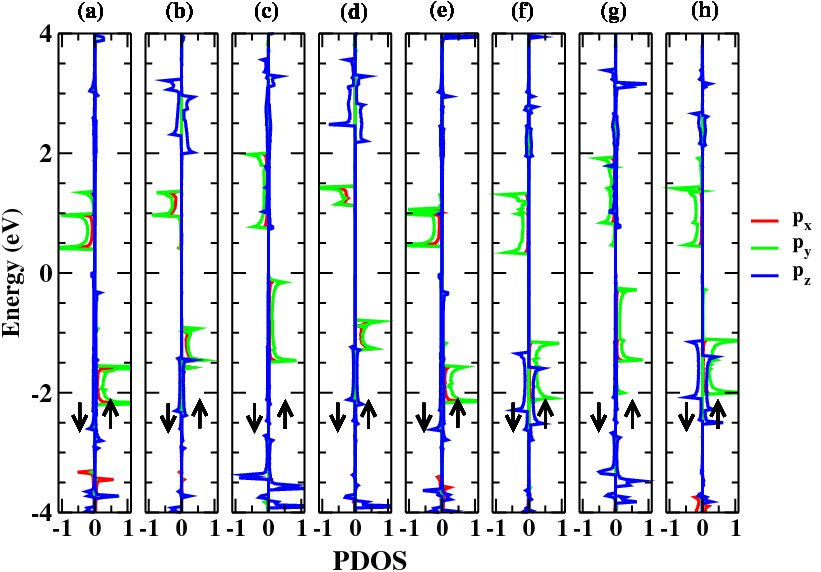}

\caption{{PDOS of p orbitals of atoms of spin-polarized 6-CC-CC-BNCNRs:
(a) upper edge bare C atom of b$_3$, (b) lower edge bare C atom of b$_3$,
(c) upper edge bare C atom of b$_4$, (d) lower edge bare C atom of b$_4$,
(e) upper edge bare C atom of b$_5$, (f) lower edge bare C atom of b$_5$,
(g) upper edge bare C atom of b$_6$, (h) lower edge bare C atom of b$_6$.
Up and down arrows denote spin-up, and spin-down, states, respectively. Configurations are explained in  the main text.}}
\end{figure}
\vspace*{-2.2in}

\begin{figure}[H]
\centering{}\includegraphics[scale=0.5]{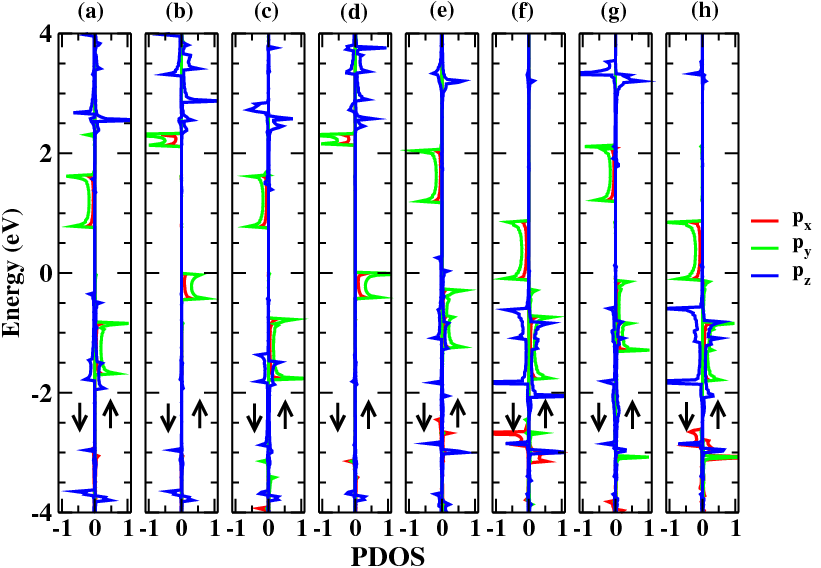}

\caption{{PDOS of p orbitals of atoms of spin-polarized 6-CC-BN-BNCNRs:
(a) upper edge bare C atom of c$_3$,
(b) lower edge bare B atom of c$_3$,
(c) upper edge bare C atom of c$_4$,
(d) lower edge bare B atom of c$_4$,
(e) upper edge bare C atom of c$_5$,
(f) lower edge bare N atom of c$_5$,
(g) upper edge bare C atom of c$_6$,
(h) lower edge bare N atom of c$_6$.
Up and down arrows denote spin-up, and spin-down, states, respectively. Configurations are explained in the main text.}}

\end{figure}

\begin{figure}[H]
\centering{}\includegraphics[scale=0.5]{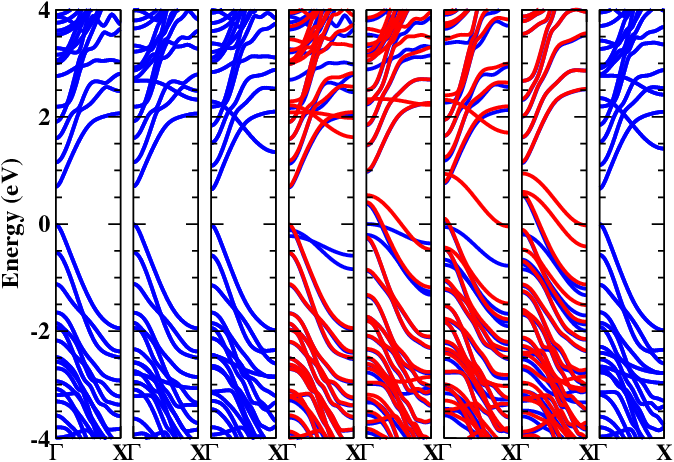}

\caption{{Band structures of various configurations of 18-BN-BN-ABNCNR:
(a) a$_0$, (b) a$_1$,
(c) a$_2$, (d) a$_3$,
(e) a$_4$, (f) a$_5$,
(g) a$_6$, and (h) a$_7$. Configurations are explained in the main text.}}

\end{figure}

\begin{figure}[H]
\centering{}\includegraphics[scale=0.5]{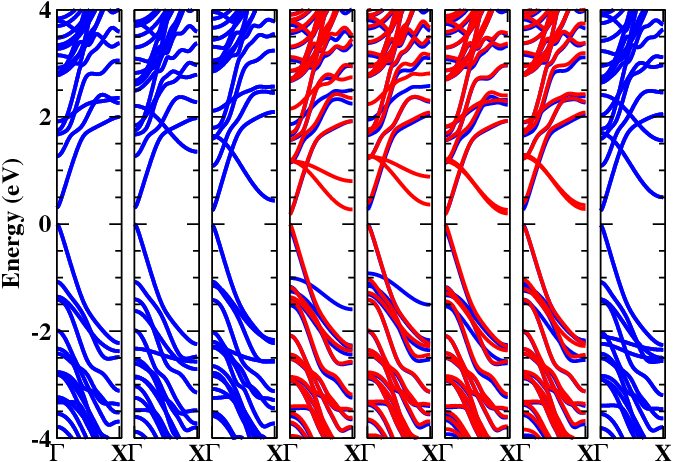}

\caption{{Band structures of various configurations of 18-CC-CC-ABNCNR:
(a) b$_0$, (b) b$_1$,
(c) b$_2$, (d) b$_3$,
(e) b$_4$, (f) b$_5$,
(g) b$_6$, and (h) b$_7$. Configurations are explained in the main text.}}

\end{figure}

\begin{figure}[H]
\centering{}\includegraphics[scale=0.5]{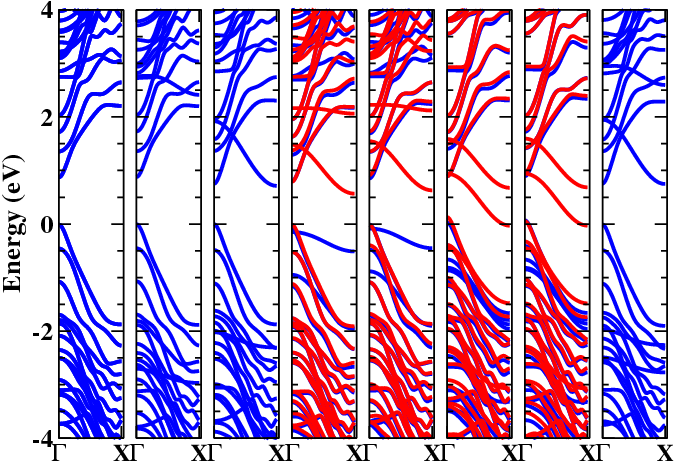}

\caption{{Band structures of various configurations of 18-CC-BN-ABNCNR:
(a) c$_0$, (b) c$_1$,
(c) c$_2$, (d) c$_3$,
(e) c$_4$, (f) c$_5$,
(g) c$_6$, and (h) c$_7$. Configurations are explained in the main text.}}
\end{figure}

\begin{table}[H]
\begin{centering}
\begin{tabular}{|c|c|c|c|c|c|c|c|c|}
\hline 
N\textsubscript{a} & \multicolumn{2}{c}{} & \multicolumn{2}{c}{Band gaps (eV)} & \multicolumn{2}{c}{} & \multicolumn{2}{c|}{}\tabularnewline
\cline{2-9} 
 & \multicolumn{2}{c|}{b\textsubscript{3}} & \multicolumn{2}{c|}{b\textsubscript{4}} & \multicolumn{2}{c|}{b\textsubscript{5}} & \multicolumn{2}{c|}{b\textsubscript{6}}\tabularnewline
\cline{2-9} 
 & Up & Dn & Up & Dn & Up & Dn & Up & Dn\tabularnewline
\hline 
8 & 1.60 & 1.07 & 1.72 & 1.15 & 1.73 & 0.86 & 1.72 & 1.05\tabularnewline
\hline 
10 & 0.86 & 0.70 & 0.85 & 0.59 & 0.87 & 0.63 & 0.92 & 0.50\tabularnewline
\hline 
12 & 0.47 & 0.24 & 0.38 & 0.60 & 0.48 & 0.23 & 0.41 & 0.33\tabularnewline
\hline 
14 & 1.17 & 0.74 & 1.14 & 0.81 & 1.17 & 0.67 & 1.16 & 0.71\tabularnewline
\hline 
16 & 1.02 & 0.62 & 0.99 & 0.72 & 1.02 & 0.53 & 1.01 & 0.63\tabularnewline
\hline 
18 & 0.33 & 0.18 & 0.27 & 0.45 & 0.34 & 0.19 & 0.27 & 0.29\tabularnewline
\hline 
\end{tabular}
\par\end{centering}
\caption{{Band gaps of CC-CC ABNNCRs for different widths. M
denotes the metallic behavior.}}
\end{table}

\begin{table}[H]
\begin{centering}
\begin{tabular}{|c|c|c|c|c|c|c|c|c|}
\hline 
N\textsubscript{a} & \multicolumn{2}{c}{} & \multicolumn{2}{c}{Band gaps (eV)} & \multicolumn{2}{c}{} & \multicolumn{2}{c|}{}\tabularnewline
\cline{2-9} 
\multirow{2}{*}{} & \multicolumn{2}{c|}{c\textsubscript{3}} & \multicolumn{2}{c|}{c\textsubscript{4}} & \multicolumn{2}{c|}{c\textsubscript{5}} & \multicolumn{1}{c}{c\textsubscript{6}} & \tabularnewline
\cline{2-9} 
 & Up & Dn & Up & Dn & Up & Dn & Up & Dn\tabularnewline
\hline 
8 & 0.50 & 0.61 & M & M & M & M & M & M\tabularnewline
\hline 
10 & 0.63 & 0.55 & M & M & M & M & M & M\tabularnewline
\hline 
12 & 0.91 & 0.94 & M & M & M & M & M & M\tabularnewline
\hline 
14 & 0.31 & 0.39 & M & M & M & M & M & M\tabularnewline
\hline 
16 & 0.67 & 0.42 & M & M & M & M & M & M\tabularnewline
\hline 
18 & 0.72 & 0.71 & M & M & M & M & M & M\tabularnewline
\hline 
\end{tabular}
\par\end{centering}
\caption{{Band gaps of BN-BN ABNNCRs for different widths. M
denotes the metallic behavior}.}
\end{table}